\documentclass[aps,onecolumn,nopacs,nofootinbib,floatfix,superscriptaddress]{revtex4}
\usepackage{graphicx}
\usepackage{amsfonts}
\usepackage{amssymb}
\usepackage{amsbsy}
\usepackage{hyperref}
\usepackage{amsmath}
\usepackage{mathrsfs}
\usepackage{latexsym}
\usepackage{natbib}
\usepackage{bm}
\usepackage{subfigure} 
\usepackage{color}
\usepackage{wasysym}
\usepackage{mathbbol}
\usepackage{bigints}
\allowdisplaybreaks
\usepackage[normalem]{ulem}
\usepackage[dvipsnames]{xcolor}
\usepackage{multirow}
\usepackage{physics}
\usepackage{comment}

\definecolor{napiergreen}{rgb}{0.16, 0.5, 0.0}

\renewcommand*{\l}{\lambda_{\star}}

\def\l{\left}
\def\r{\right}

\def\cosech{\textup{cosech}}

\begin{document}

\title{Timelike transitions in an atom by a mirror in light cone and Kruskal-Szekeres regions: a status of quantum equivalence}

\author{Subhajit Barman}
\email{subhajit.barman@physics.iitm.ac.in}
\affiliation{Centre for Strings, Gravitation and Cosmology, Department of Physics, Indian Institute of Technology Madras, Chennai 600036, India}

\author{Pradeep Kumar Kumawat}
\email{pradeep.kumawat@iitg.ac.in}
\affiliation{Department of Physics, Indian Institute of Technology Guwahati, Guwahati 781039, Assam, India}

\author{Bibhas Ranjan Majhi}
\email{bibhas.majhi@iitg.ac.in}
\affiliation{Department of Physics, Indian Institute of Technology Guwahati, Guwahati 781039, Assam, India}

\begin{abstract}
We investigate the timelike transitions in a two-level atom in the presence of an infinite reflecting mirror in the future-past light cone regions of a Minkowski spacetime, as well as in the region interior of a $(1+1)$ dimensional Schwarzschild black hole. In particular, when considering the light cone regions, two specific scenarios are dealt with -- $(i)$ a mirror is static in Minkowski spacetime while the atom is attached to a frame confined inside the future or past light cone region, $(ii)$ an atom is static in Minkowski spacetime, and the mirror is confined inside the future or past light cone region. For both situations the atom is interacting with field modes defined in mirror's frame. Analogous configurations are considered in the black hole spacetime: in one case, the mirror carries field modes represented by the Kruskal time, while the atom follows the Schwarzschild time defined inside the black hole; in the other case, the situations are reversed. The analyses, depending upon the frame of the atom, are respectively done within the light cone, Minkowski, Schwarzschild, and Kruskal time-interaction pictures. In all of these scenarios, we observe that the excitation probabilities contain a thermal factor and depend periodically on the separation between the atom and the mirror. At the level of transition probabilities, the aforesaid two scenarios in $(1+1)$ dimensional Minkowski-light cone regions appear to be the same for the equal field and atomic frequencies. However, the same is not true when we consider the $(3+1)$ dimensional Minkowski-light cone or the Schwarzschild interior regions. We also estimate the de-excitation probabilities and encounter similar situations. 
However, we observe that the excitation to de-excitation ratios (EDRs) corresponding to analogous scenarios are equal for equal atomic and field frequencies. This bolsters our earlier observation on the relevance of EDRs \cite{Kumawat:2024kul} in the context of the quantum equivalence between analogous scenarios.

\end{abstract}

\date{\today}

\maketitle
 
\section{Introduction}\label{sec:Introduction}

Einstein's equivalence principle (EEP) \cite{Paunkovic:2022flx} -- the trajectories of a particle under gravity are indistinguishable locally from those for a free particle viewed with respect to an accelerated frame -- motivates the study of physics in the accelerated frame to understand the local features of gravity. In fact, as a consequence of this equivalence, one finds that an accelerated frame (inertial frame) in Minkowski spacetime is comparable to a local stationary (freely falling) frame in a black hole spacetime. Although the validity of these equivalences at the quantum level is sometimes questionable (see \cite{Pipa:2018bui, Paunkovic:2022flx, Zych:2015fka, Viola:1996de} for various investigations), there have been several works that also support the existence of this equivalence \cite{Svidzinsky:2018jkp, Fulling:2018lez, Lammerzahl:1996se} or propose experimental tests for it \cite{Geiger:2018xwr, Tino:2020nla}. The question of the validity of the equivalence principle at the quantum level puts forward another aspect related to the equivalent motions seen from Rindler and Minkowski frames. Interestingly, in \cite{Svidzinsky:2018jkp}, the quantum equivalence between these two observers is presented. It is to be noted that though these observers seem to be in different classical motions, there is a symmetry in their relative motions -- the motion of a uniformly accelerating observer seen from the Minkowski frame as compared to a Minkowski observer seen from the accelerated frame.

In particular, in \cite{Svidzinsky:2018jkp}, atomic excitations in the presence of an infinite mirror are analyzed between two distinct scenarios to understand this quantum equivalence. Here a two-level atom interacts with field modes, described in the mirror frame. The mirror mimics a boundary in the spacetime and the field modes are obtained using the Dirichlet boundary condition (see \cite{book:Birrell}). 
The mirrors are useful tools to understand phenomena on static and noninertial trajectories \cite{Foo:2020rsy, Brown:2015yma, WanMokhtar:2022npv} as well as dynamical Casimir effect~\cite{Hawking:1979ig, Moore:1970tmc, Davies:1977yv, Davies:1976hi, Hotta:2015yla, Good:2016atu, Good:2018ell, Fernandez-Silvestre:2021ghq, Myrzakul:2021bgj, PhysRevD.94.065010, Zhou:2013hx}. 
In \cite{Svidzinsky:2018jkp}, two scenarios are considered -- Case (I): a mirror is static in Minkowski spacetime while the atom is uniformly accelerating, and Case (II): the mirror is accelerating uniformly while the atom is static in the Minkowski frame. The investigation in \cite{Svidzinsky:2018jkp} concludes that the transition probabilities of the atom corresponding to these two scenarios are identical when the photon and atom frequencies are the same. It is to be noted that in their work, scalar field theory is considered, which is a good approximation for the electromagnetic theory in the absence of angular momentum exchange, and thus the scalar theory is referred in terms of photons. Subsequently, in this work, the quantum equivalence between observers in Minkowski and black hole spacetimes is also indicated. However, the analysis of \cite{Svidzinsky:2018jkp} is done in $(1+1)$ dimensions and in $(3+1)$ dimensions the same equivalence between Case (I) and Case (II) may not remain valid, which is shown in \cite{Kumawat:2024kul}. Moreover, it also seems necessary to provide a field mode independent realization of the above equivalence, as in \cite{Svidzinsky:2018jkp} the analysis is done considering a single frequency mode.\vspace{0.1cm}

\begin{figure}[h!]
\centering
\includegraphics[width=17cm]{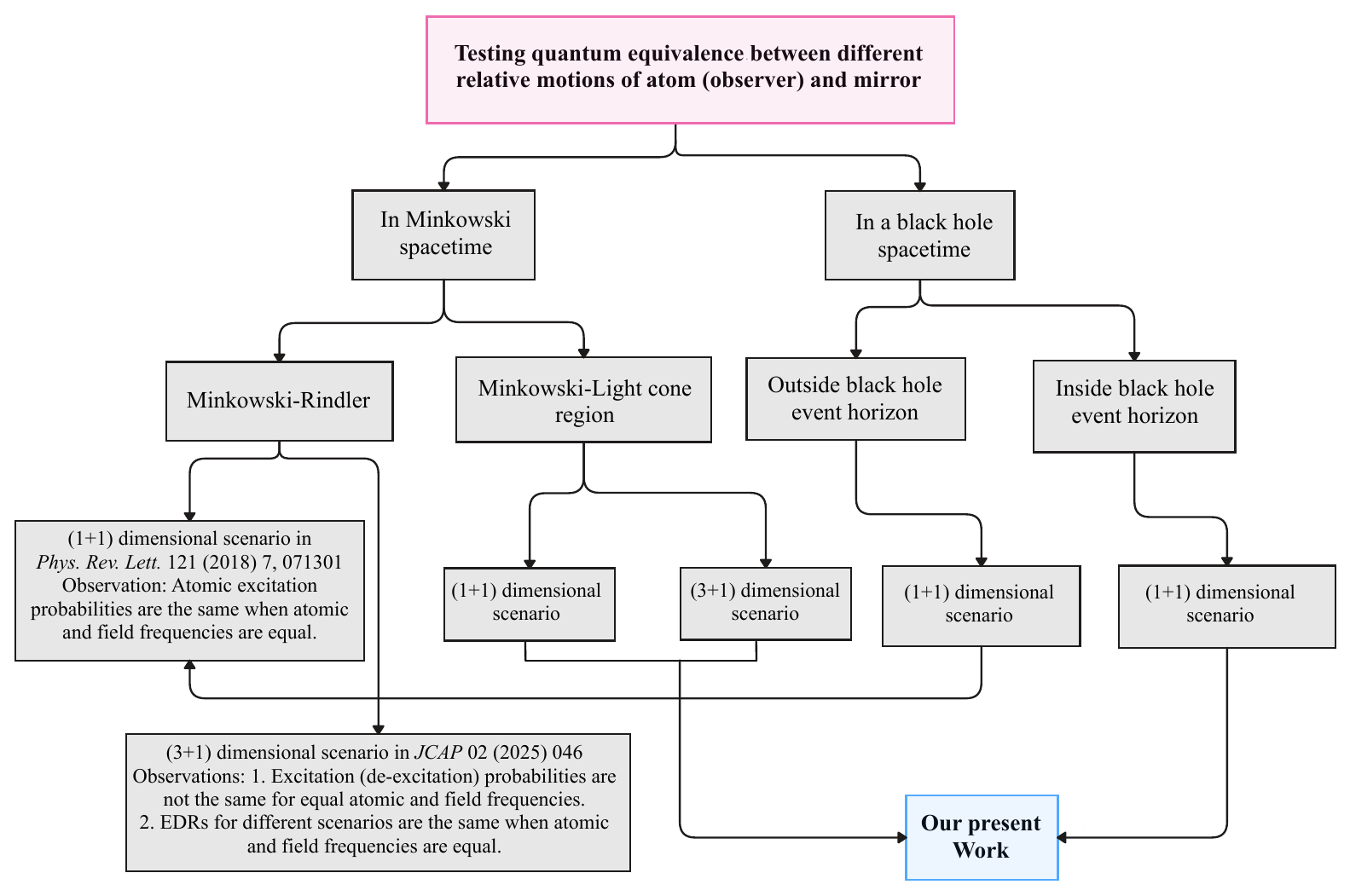}
\caption{The above diagram depicts the different areas that have been or are being studied in regard to understanding the quantum equivalence between different scenarios with atom-mirror set-ups. Our present work comprises set-ups prepared in the Minkowski-Light cone regions, and also compares the results with set-ups inside the event horizon of a Schwarzschild black hole.}
\label{fig:Quantum-Eqv}
\end{figure}

Hence, we feel this apparent symmetry at the quantum level between atoms in different classical motions needs to be further investigated with care. Also, in order to check the width of validity of this type of quantum equivalences, it is necessary to extend the investigations to different classical motions. Hence, our aims in this paper are twofold. Firstly, we would like to investigate whether the observed equivalence in \cite{Svidzinsky:2018jkp} for $(1+1)$ dimensions holds for other types of relative motions. Secondly, it will be interesting to know the situations in $(3+1)$ dimensions.
For these objectives, we will check the symmetry at the quantum level when there is symmetry in the relative time progress between two frames. In this regard, we consider the Minkowski spacetime's future and past light cone regions and compare them with an observer following the Minkowski coordinates, see \cite{Olson:2010jy, Quach:2021vzo}. The particle creation observed by a quantum probe confined inside these light cone regions is termed the `Timelike Unruh Effect'. One can check Fig. \ref{fig:Quantum-Eqv} for a diagrammatic representation of the areas that are being investigated in the study of quantum equivalence with atom-mirror set-ups. We have also specified the areas that our present work covers.

We consider two frames -- one for a quantum system, taken as a two-level atom, and another one for a mirror. The atom is interacting with the field modes in the mirror's frame. With this, we investigate the excitation probability of the atom in two different scenarios. 
\begin{itemize}
\item Case $(i)$: In the first scenario, the mirror is kept static at $z = z_0$ in the Minkowski spacetime described by the coordinates $(t, \,z)$, and thus the field modes evolve with time $t$. At the same time, the atom is confined in the future light cone region with coordinates $(\eta, \zeta)$, at a fixed $\zeta = 0$. Therefore, the atom's evolution is recorded with respect to the light cone time $\eta$.

\item Case $(ii)$: In the second scenario, the atom is static at $z = z_0$ in the Minkowski spacetime, with its proper time associated with Minkowski time $t$. At the same time, the mirror is confined in the light cone region at $\zeta = 0$, and thus the field modes evolve with the light cone time $\eta$. It is to be noted that in Case $(ii)$ the atom and mirror positions are interchanged as compared to Case $(i)$.
\end{itemize}
The above analysis within the light cone regions can be important for the Milne universe. This is because the light cone region is analogous to the geometry of the Milne Universe, which corresponds to a special scenario in the Friedmann–Lema\^itre–Robertson–Walker (FLRW) metric \cite{book:Birrell, mukhanov2005physical}. 
Moreover, the possibility of imitating the Milne universe in an analogous gravitational system \cite{Sanchez-Kuntz:2022gds}, provides additional relevance to the current investigation. 
Our observation, in the context of light cone regions, suggests that the structures of the single photon transition probabilities corresponding to different scenarios are equivalent in $(1+1)$ dimensions when the atomic and field frequencies are equal, while the same is not true in $(3+1)$-spacetime dimensions.
The same situation also arises for the Minkowski-Rindler case, which is discussed thoroughly in \cite{Kumawat:2024kul}.\vspace{0.1cm}

For a better understanding of the first part of our aforesaid two-fold aims, we also consider set-ups inside the event horizon of a Schwarzschild black hole. Note that the Kruskal-Szekeres (KS) coordinate transformation inside a black hole is analogous to the coordinate transformation to the light cone region. Our investigation of atomic transitions and related equivalence between the light cone and Schwarzschild interior regions should indicate physical insights that are consequences of Einstein's equivalence-like principle (EELP)\footnote{Note that Einstein's equivalence is strictly reserved for accelerated motion. Therefore, the present detector motion in the light cone regions and their analogy with the Schwarzschild interior are termed differently, and we prefer to call it the consequences of Einstein's equivalence-like principle (EELP).}.
Here we consider two specific scenarios. 
\begin{itemize}
    \item In Case~$(i)$, the mirror frame, fixed at $X = X_0$, carries field modes represented in Kruskal--Szekeres (KS) coordinates $(X, T)$. At the same time, the atom uses Schwarzschild coordinates $(t_{S}, r_{\star})$, defined inside the black hole, where $t_{S}$ is a spacelike coordinate and $r_{\star}$ (tortoise coordinate) is a timelike coordinate. The atom follows a particular trajectory along which $t_{S}$ remains constant ($t_{S} = 0$), whereas $r_{\star}$ is changing, and so it can move toward the singularity at $r = 0$.

    \item In Case~$(ii)$, the coordinate systems are interchanged. Now, the atom uses KS coordinates.  It is fixed at $X = X_0$ and it evolves with KS time $T$. At the same time, the mirror frame carries field modes represented in Schwarzschild coordinates, defined inside the black hole. Consequently, the mirror follows the $t_{S}=0$ trajectory.
\end{itemize}
Our observations suggest that even in two spacetime dimensions, atomic transition probabilities corresponding to different scenarios in the Schwarzschild interior region do not show the aforesaid equivalence when the atomic and field frequencies are equal. This, however, does not happen outside the black hole horizon \cite{Svidzinsky:2018jkp} and is also in contrast with ($1+1$)-dimensional Minkowski-light cone results. Furthermore, these atomic transitions from the Schwarzschild interior are not equivalent to those from the Minkowski-light cone regions. We believe these dissimilarities in atomic transitions between the Schwarzschild interior and Minkowski-light cone regions, as compared to the Rindler-Minkowski and outside BH-KS correspondence, can provide us with crucial insight regarding the effect of spacetime geometry on atomic transitions. Overall, our observations suggest that at the level of atomic transition probabilities, the EELP is not satisfied.\vspace{0.1cm}

Further, in \cite{Kumawat:2024kul}, we have seen that though the atomic excitation or de-excitation probabilities between classically equivalent scenarios are not identical, the excitation to de-excitation ratios (EDRs) can match. Therefore, EDR can be a relevant quantity to study the equivalence principle at the quantum level. Our observations suggest that in $(3+1)$ dimensions corresponding to the Minkowski-light cone region, the EDRs are the same when the atomic and field frequencies are equal. At the same time, for the Schwarzschild interior, when we are comparing between the cases -- (i) the atom is fixed inside the black hole with respect to the spacelike Schwarzschild coordinate $t$ and (ii) the atom is fixed with respect to the Kruskal spacelike coordinate -- the EDRs are equal only in the near-horizon regime and for certain conditions. However, we would like to mention that the EDRs corresponding to the two different backgrounds, i.e., Minkowski-light cone and BH interior, do not match. Therefore, from the perspective of EDRs the EELP is still not valid at the quantum level.\vspace{0.1cm}

This manuscript is organized in the following manner. In Sec. \ref{sec:Coords}, we provide the coordinate transformations to the light cone regions in a Minkowski background, and recall the Kruskal coordinates inside a Schwarzschild black hole event horizon. In this section, we also elucidate the set-up for investigating the transition probability of a two-level atom. In Sec. \ref{sec:Detector-response}, we investigate the excitation probabilities of these atoms in the presence of a mirror in $(1+1)$ and $(3+1)$ dimensional light cone regions. Subsequently, in Sec. \ref{sec:Sch-transition}, we investigate the excitation of atoms inside the event horizon of a Schwarzschild black hole. In Sec. \ref{sec:de-excitation}, we study the de-excitation probabilities of the atoms for each of these scenarios and estimate the EDRs. We discuss our observations and specify the resemblance between the scenarios of Secs. \ref{sec:Detector-response} and \ref{sec:Sch-transition}, and provide the concluding remarks in Sec. \ref{sec:discussion}.

\section{Coordinate transformations and the set-up}\label{sec:Coords}
In this section, we specify the coordinate transformations suitable for the future and the past light cone regions in a Minkowski background.  We also recall the Kruskal-Szekeres (KS) coordinate transformation suitable for a freely falling observer inside the event horizon of a Schwarzschild black hole. It is to be noted that all of these coordinate transformations will be presented for the $(1+1)$ dimensional scenario, and their representation in $(3+1)$ dimensions follows readily. In this section, we will also present the setup necessary to understand the transition of two-level atoms interacting with the background scalar field.

\subsection{Coordinate transformations}
Let us first recall sets of coordinates suitable for the future or the past light cone regions of a $(1+1)$ dimensional Minkowski spacetime (see \cite{Crispino:2007eb}). In particular, for an observer confined in the future light cone region \cite{Olson:2010jy, Quach:2021vzo}, the coordinate relations are
\begin{subequations}\label{eq:CT-F}
\begin{eqnarray}
    t &=& \frac{c}{a}\,e^{a\,\eta/c} \cosh{(a\,\zeta/c^2)}~;\\
    ~z &=& \frac{c^2}{a}\,e^{a\,\eta/c} \sinh{(a\,\zeta/c^2)}~,
\end{eqnarray}
\end{subequations}
where
\begin{figure}[h!]
\centering
\includegraphics[width=7.5cm,height=7.4cm]{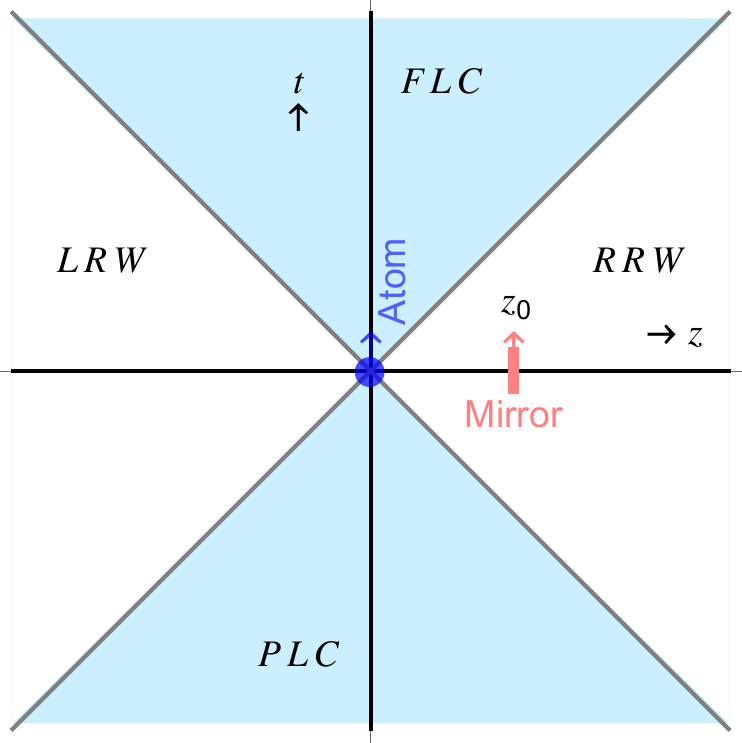}
\hskip 30pt 
\includegraphics[width=7.5cm,height=7.4cm]{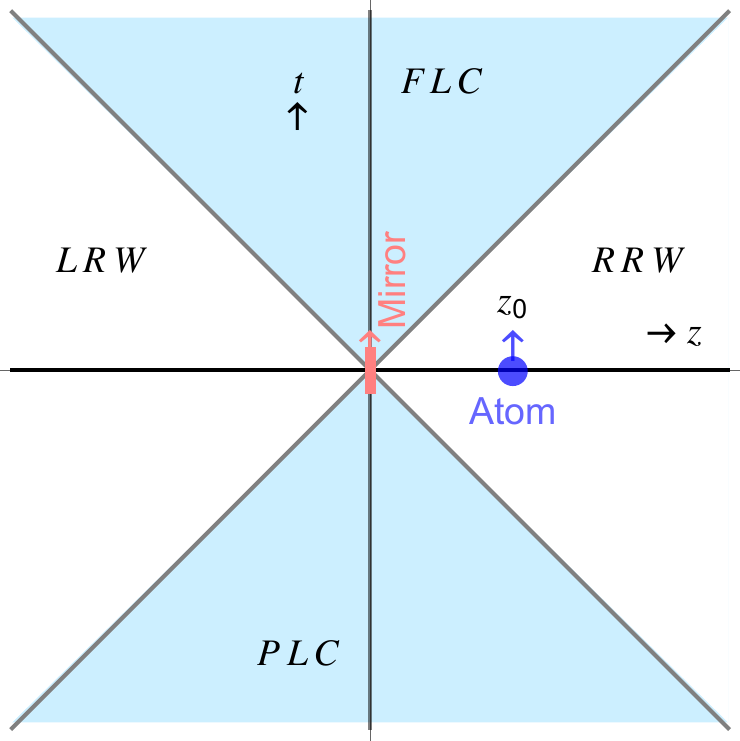}
\caption{The above figures provide schematic diagrams depicting different regions in the Minkowski spacetime. The region $RRW$ denotes the right Rindler wedge, while $LRW$ denotes the left Rindler wedge. The top and bottom shaded regions, respectively, denote the future and the past light cone regions ($FLC$ and $PLC$). The above two figures correspond to two different atom and mirror positions. On the left, we have depicted a scenario where the Mirror is kept fixed at $z=z_{0}$, and the atom is in the region $FLC$ at $z=0$. On the right, the mirror is at $z=0$, and the atom is kept fixed at $z=z_{0}$.}
\label{fig:SD-Kasner}
\end{figure}
$(t,z)$ are the Minkowski coordinates, and $(\eta,\zeta)$ are the coordinate choice for future light cone ($FLC$). It is to be noted that in the above expression $c$ corresponds to the speed of light in vacuum, and $a$ is another real-positive parameter which has the dimension of acceleration. Therefore, $a/c$ has the dimension of inverse time or frequency. One can notice that the above coordinate transformation \eqref{fig:SD-Kasner} is bounded in the future region. This is evident from the constraint $c^2\,t^2-z^2=c^4\,a^{-2}\,e^{2a\,\eta/c}$, which indicates to $t>0$. 
We would also like to mention that for an observer at $z=0$ in the light cone region, the proper time $\tau$ is related to the conformal time $\eta$ via the relation $\tau=(c/a)\,e^{a\eta/c}$. Thus, the parameter $a$, more specifically $a/c$, acts as a scaling between the proper and conformal times. In fact, the coordinate transformation of Eq. \eqref{eq:CT-F} results in a metric that is analogous to the Milne Universe (see \cite{book:Birrell}), where $a$ mimics the scale factor.

The coordinates $(\bar{\eta},\bar{\zeta})$ suitable for an observer in the past light cone ($PLC$) are related to the Minkowski coordinates $(t,z)$ as
\begin{subequations}\label{eq:CT-P}
\begin{eqnarray}
    t &=& -\frac{c}{a}\,e^{a\,\bar{\eta}/c} \cosh{(a\,\bar{\zeta}/c^2)}~;\\
    ~z &=& -\frac{c^2}{a}\,e^{a\,\bar{\eta}/c} \sinh{(a\,\bar{\zeta}/c^2)}~.
\end{eqnarray}
\end{subequations}
In contrast to the previous case, in this scenario $t<0$, and the coordinate transformation corresponds to an observer confined in the past. We would like to mention that both of these above regions of $FLC$ and $PLC$ are depicted in schematic diagrams in Fig. \ref{fig:SD-Kasner}. We would also like to highlight the analogy between the geometry of light cone regions and a special scenario of the Friedmann–Lema\^itre–Robertson–Walker (FLRW) spacetime. This scenario, known as the Milne universe, corresponds to the scale factor given by the coordinate time of the FLRW spacetime, see \cite{book:Birrell,book:Mukhanov}. The Milne universe can correspond to expanding cosmological models with zero spacetime curvature. Actually, the light cone-Minkowski type coordinate description also covers a part of the Milne Universe. Thus, with this analogy, the observations related to light cone regions will be applicable to a part of the Milne Universe as well.

\begin{figure}[h!]
\centering
\includegraphics[width=7.5cm,height=7.4cm]{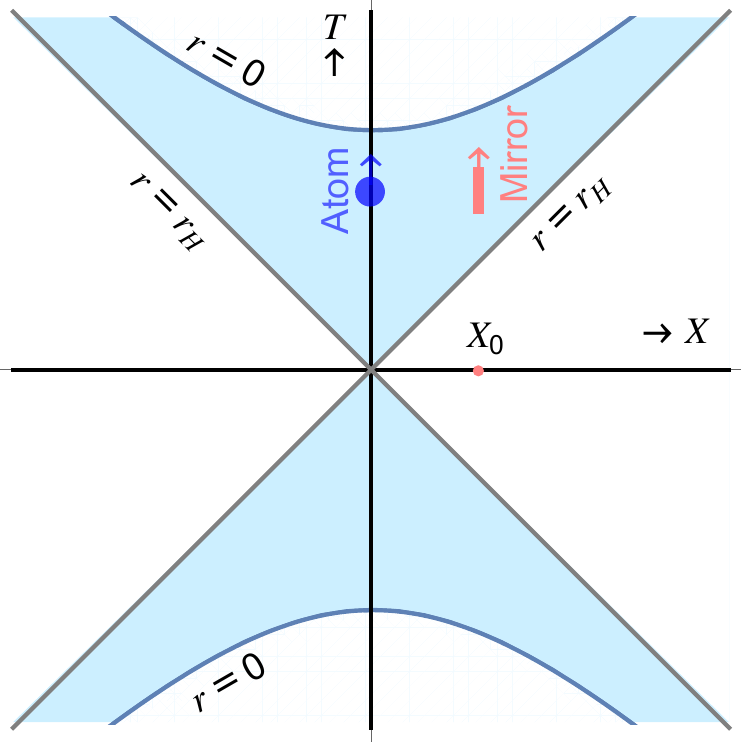}
\hskip 30pt 
\includegraphics[width=7.5cm,height=7.4cm]{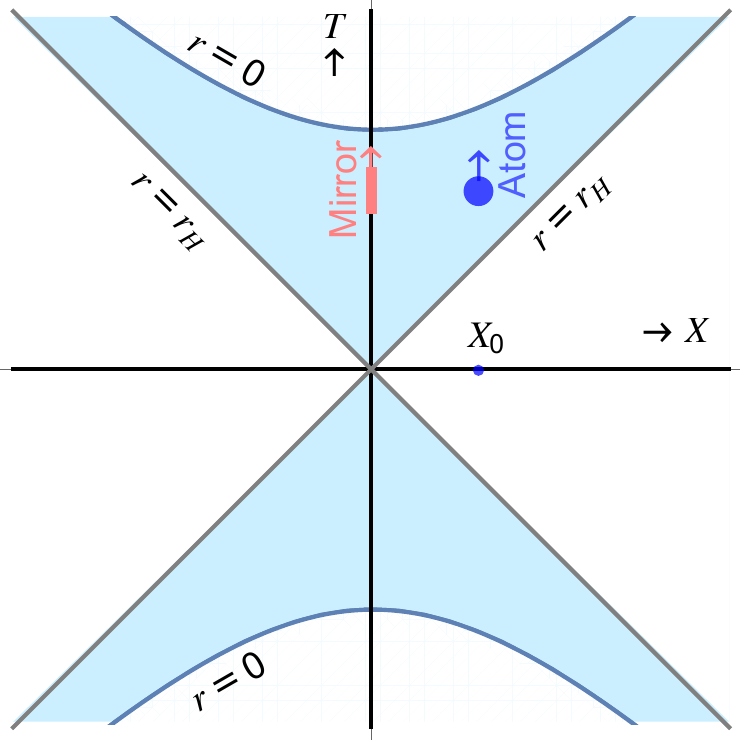}
\caption{The above figures provide schematic diagrams depicting the Kruskal–Szekeres representation of the Schwarzschild black hole spacetime. The $45^{\circ}$ grey lines denote the event horizon $(r=r_{H})$. The $r=0$ singularity is depicted by curved lines. The shaded regions correspond to the interior of the black hole. The above two figures correspond to two different set-ups for the atom-mirror positions. For instance, on the left, the mirror is free-falling, described by a fixed KS space coordinate $X=X_{0}$. In contrast, on the right, the atom is free-falling, described by $X=X_{0}$.}
\label{fig:SD-Kruskal}
\end{figure}
Next, we consider a Schwarzschild black hole spacetime, and we will present the Kruskal coordinate transformation suitable for the region inside its event horizon. In $(1+1)$ dimensions, the line element in this background, with coordinates $(t_{S},r)$, is given by
\begin{eqnarray}\label{eq:SchIn-Metric}
    ds^2 &=& -\l(1-\frac{r_{H}}{r}\r)\,c^2\,dt_{S}^2+\l(1-\frac{r_{H}}{r}\r)^{-1}\,dr^2~\nonumber\\
    ~&=& \l(1-\frac{r_{H}}{r}\r)\,\Big[-c^2\,dt_{S}^2+dr^2_{\star}\,\Big]~,
\end{eqnarray}
where $r_{H}$ denotes the position of the event horizon from the central singularity, and $r_{\star}$ is the tortoise coordinate defined as $dr_{\star}=(1-r_{H}/r)^{-1}dr$. The KS coordinates suitable for the region inside the black hole, see \cite{Padmanabhan:2010zzb,Banerjee:2009wb}, are given by 
\begin{eqnarray}\label{eq:SchIn-Kruskal}
    T = \frac{1}{c\,\kappa_{H}}\,e^{\kappa_{H}\,r_{\star}}\,\cosh{(c\,\kappa_{H}\,t_{S})}~,~~X = \frac{1}{\kappa_{H}}\,e^{\kappa_{H}\,r_{\star}}\,\sinh{(c\,\kappa_{H}\,t_{S})}~;
\end{eqnarray}
where, $\kappa_{H}=1/(2\,r_{H})$ is the surface gravity at the horizon. Furthermore, we consider redefined coordinates inside the black hole, namely the advanced and the retarded time coordinates, $v=r_{\star}+c\,t_{S}$ and $u=r_{\star}-c\,t_{S}$. Then one can obtain the relation between these coordinates and the KS coordinates as 
\begin{eqnarray}\label{eq:Sch-Kruskal-EF}
c\,T+X= \frac{e^{\kappa_{H}\,v}}{\kappa_{H}}; \,\,\,\,\ c\,T-X= \frac{e^{\kappa_{H}\,u}}{\kappa_{H}}~.
\end{eqnarray}
The coordinate transformations mentioned above will be useful when estimating the atomic transition probabilities in different scenarios. The KS diagram for a Schwarzschild black hole, and specifically the black hole interior region in this diagram, is depicted in Fig. \ref{fig:SD-Kruskal}.

\subsection{Response of atom: the set-up}

Here, we elucidate the set-up for investigating the transition probability of electrically neutral atoms. In particular, we consider an atom interacting with a scalar field $\phi$ in the considered background spacetime, be that the light cone regions in Minkowski or the future Kruskal region of the Schwarzschild. We consider the atoms to be two-level systems. Depending on different backgrounds and atom trajectories, the set-ups to detect the atomic transition will have subtle changes. For instance, when the atom is in the light cone region of the Minkowski spacetime, it is convenient to consider the atomic Hamiltonian to correspond to the conformal time $\eta$, see \eqref{eq:CT-F}, which correctly indicates the dynamical nature of the coordinate transformation described by that region, see \cite{Olson:2010jy, Quach:2021vzo}. Similarly, for other scenarios, one should also express the atomic and interaction  Hamiltonian in coordinates that best represent the background and the trajectories. In the following analysis, we obtain these Hamiltonians and the corresponding formulas for the atomic transition probabilities.

\subsubsection{Atom confined in light cone region of the Minkowski spacetime}

Here, we consider the atom confined inside the future light region of the Minkowski spacetime. To elucidate the above proposition let us consider the initial atom-field state to be $|\Psi\rangle$ and the Hamiltonian is $\hat{H} = \hat{H}_{A} + \hat{H}_{F} + \hat{H}_{I}$, where $\hat{H}_{A}$, $\hat{H}_{F}$, and $\hat{H}_{I}$ denote the free-atomic, free-field, and the interaction Hamiltonian respectively. In particular, this Hamiltonian is best expressed in terms of the atomic proper time $\tau$, i.e., with respect to a clock that is comoving with the atom. In this comoving frame, the atomic energy levels do not contain any effects due to their motion. However, a question remains -- what will an observer see if it is not comoving with the atom? In that scenario, one needs to understand the change in the Hamiltonian corresponding to that specific trajectory. In particular, the Schrodinger equation with respect to time $\eta$ will be given by $i\,\partial|\Psi\rangle/ \partial \eta = (\partial\tau/ \partial\eta) \hat{H}|\Psi\rangle$, see \cite{Olson:2010jy}. This motivates us to deduce that with respect to time $\eta$, the interaction Hamiltonian is modified by a factor of $(\partial\tau/\partial\eta)$. For an atom in the light cone region of the Minkowski spacetime, we consider $\omega_{\eta}$ to be the angular frequency corresponding to the atomic energy gap, which in turn corresponds to the conformal time $\eta$, i.e., $\omega_{\eta}$ is obtained from the redshifted atomic Hamiltonian $(\partial\tau/ \partial\eta) \hat{H}_{A}$. We consider the Hamiltonian $\hat{H}_{I}(\eta)$ to denote the interaction between the field and the atom, and realized with respect to time $\eta$. This interaction Hamiltonian is considered to be linear in the field operator of arbitrary frequency $\nu$ \cite{Svidzinsky:2018jkp}, an assumption motivated by the earlier works on light-matter interaction \cite{Dicke:1954}. It is to be noted that as this Hamiltonian corresponds to time $\eta$, it should be appropriately redshifted by a multiplicative factor $\partial\tau/ \partial\eta$, see \cite{Olson:2010jy}. From Eq. \eqref{eq:CT-F} we observe that this redshift factor $\partial\tau/ \partial\eta = e^{a\,\eta/c}$, in the future light cone region. This interaction Hamiltonian $\hat{H}_{I}(\eta)$ corresponding to time $\eta$ can then be explicitly expressed as
\begin{eqnarray}\label{eq:Int-Hamiltonian-kasner1}
    \hat{H}_{I}(\eta) &=& \hbar\,g\,e^{a\,\eta/c}\,\Big(\phi_{\nu}[t(\eta),z(\eta)]\,\hat{a}_{\nu}+h.c.\Big)\,\big(e^{-i\,\omega_{\eta}\,\eta}\,\hat{\sigma}+h.c.\big)~.
\end{eqnarray}
In the above expression, $\hbar$ is the Planck constant, $g$ is the atom-field interaction strength, $\hat{a}_{\nu}$ is the scalar field annihilation operator with frequency $\nu$, $\hat{\sigma}$ is the atomic lowering operator, and $\phi_{\nu}$ denotes the scalar field mode function.
One can notice that the above interaction Hamiltonian is modified by a nontrivial factor $\partial\tau/\partial\eta=e^{a\,\eta/c}$, as compared to its somewhat similar definition with respect to the proper time $\tau$ \cite{Svidzinsky:2018jkp}. Let the states $|\omega_{0}\rangle$ and $|\omega_{1}\rangle$ respectively denote the ground and the excited atomic states. We also consider that the field is initially prepared in the vacuum state $|0\rangle$. We want to find out the probability for the atom to get excited and simultaneously emit a single photon. In particular, this excitation probability corresponding to the atom of frequency $\omega_{\eta}$, see \cite{Svidzinsky:2018jkp}, will be
\begin{eqnarray}\label{eq:TP-gen-kasner1}
    \mathcal{P}^{ex}_{\nu} (\omega_{\eta}) &=& \frac{1}{\hbar^2} \bigg|\int d\eta\, \langle 1_{\nu},\omega_{1}|\hat{H}_{I}(\eta)|0,\omega_{0}\rangle\bigg|^2~\nonumber\\
    ~&=& g^2\bigg|\int_{-\infty}^{\infty} d\eta\,e^{a\,\eta/c}\,\phi^{\star}_{\nu}\big[t(\eta),z(\eta)\big]\,e^{i\,\omega_{\eta}\,\eta}\bigg|^2~.
\end{eqnarray}
To obtain the above expression for the transition probability, we have considered the leading non-trivial order in $g\ll 1$ term in the perturbation series. One should note that the above set-up to investigate the transition probability is specific to an atom in the future light cone region of the Minkowski spacetime.

\subsubsection{Atom fixed with the Minkowski frame}

If the atom is in the Minkowski spacetime, the proper time $\tau$ is given by $t$. In this scenario, we consider the atomic angular frequency corresponding to the energy gap to be $\omega_{t}$. Moreover, here the redshift factor is precisely unity. Then the interaction Hamiltonian and the atomic excitation probability can be expressed as 
\begin{subequations}
\begin{eqnarray}\label{eq:Int-Hamiltonian-kasner2}
    \hat{H}_{I}(t) &=& \hbar\,g\,\Big(\phi_{\nu}[t,z]\,\hat{a}_{\nu}+h.c.\Big)\,\big(e^{-i\,\omega_{t}\,t}\,\hat{\sigma}+h.c.\big)~;\\
    \mathcal{P}^{ex}_{\nu} (\omega_{t}) &=& g^2\bigg|\int_{-\infty}^{\infty} dt\,\phi^{\star}_{\nu}\big[t,z\big]\,e^{i\,\omega_{t}\,t}\bigg|^2~.
    \label{eq:TP-gen-kasner2}
\end{eqnarray}
\end{subequations}
It is to be noted that this expression corresponds to an atom that has the proper time $t$. This particular set-up was initially proposed in \cite{Svidzinsky:2018jkp} and also used in \cite{Chakraborty:2019ltu}.

\subsubsection{Atom inside Schwarzschild event horizon and following tortoise coordinates}

Next, if the atom were to be inside the event horizon of the Schwarzschild spacetime, see Eq. \eqref{eq:SchIn-Metric}, the conformal time described by the coordinate $(r_{\star}/c)$ is appropriate to capture the specific essence of the region. In this scenario, we consider the atomic angular frequency corresponding to the atomic energy gap that is tuned with the conformal time $(r_{\star}/c)$ to be $\omega_{r_{\star}}$. Moreover, we consider $\tau$ to be the proper time in this scenario. Then with the help of Eq. \eqref{eq:SchIn-Metric} we can find the redshift factor to be $c\,\partial\tau/\partial r_{\star}=(-1+r_{H}/r)^{1/2}$. Near the event horizon $r=r_{H}$, this redshift factor can be approximated to be $c\,\partial\tau/\partial r_{\star} \simeq e^{r_{\star}/(2\,r_{H})} = e^{\kappa_{H}\,r_{\star}}$. With the help of this redshift factor, one can express the interaction Hamiltonian and the atomic excitation probabilities inside and near the event horizon in an $(1+1)$ dimensional Schwarzschild black hole spacetime as
\begin{subequations}
\begin{eqnarray}\label{eq:Int-Hamiltonian-Sch1}
    \hat{H}_{I}(r_{\star}) &=& \hbar\,g\,c\,\frac{\partial\tau}{\partial r_{\star}}\,\Big(\phi_{\nu}\big[t(r_{\star}),z(r_{\star})\big]\,\hat{a}_{\nu}+h.c.\Big)\,\big(e^{-i\,\omega_{r_{\star}}r_{\star}/c}\,\hat{\sigma}+h.c.\big)~;\\
    \mathcal{P}^{ex}_{\nu} (\omega_{r_{\star}}) &=& g^2\bigg|\int_{-\infty}^{r_{\star}^{f}} \frac{dr_{\star}}{c}\,e^{\kappa_{H}\,r_{\star}}\,\phi^{\star}_{\nu}\big[t(r_{\star}),z(r_{\star})\big]\,e^{i\,\omega_{r_{\star}}r_{\star}/c}\bigg|^2~.\label{eq:TP-gen-Sch1}
\end{eqnarray}
\end{subequations}
In the above expressions, the constant $c$ arrives from the definition of the conformal time, which is $(r_{\star}/c)$. We would like to mention that in the above expression, we truncated the integration over $r_{\star}$ at $r_{\star}^{f}$, as we are interested in observations near the horizon and this expression can be obtained analytically (here, \( r_{\star}^{f} \) is a finite cutoff inside the horizon, but not extended all the way to the singularity). With the general expression of $\partial\tau/\partial r_{\star}$, this integration should have been carried out up to $r_{\star}=0$, which corresponds to the Schwarzschild central singularity. In this second scenario, with a general $\partial\tau/\partial r_{\star}$ inside a black hole, we could not obtain the results analytically, which also contributes to another reason why we chose to obtain the transition probability near the horizon.

\subsubsection{Atom inside Schwarzschild event horizon and following Kruskal coordinates}

With the help of Eqs. \eqref{eq:SchIn-Metric} and \eqref{eq:SchIn-Kruskal}, one can notice that the Schwarzschild line element in terms of the Kruskal coordinates reads as $ds^2=(1-r_{H}/r)\,e^{-2\,\kappa_{H}r_{\star}}[c^2\,dT^2-dX^2]$. Thus inside the event horizon of an $(1+1)$ dimensional Schwarzschild black hole the Kruskal coordinate-time $T$ and the proper time $\tau$ are related among themselves as $\partial\tau/\partial T=(-1+r_{H}/r)^{1/2}\,e^{-\,\kappa_{H}r_{\star}}$. For such an atom following the trajectory described by the Kruskal coordinate transformation \eqref{eq:SchIn-Kruskal}, the interaction Hamiltonian and the atomic excitation probability are given by
\begin{subequations}
\begin{eqnarray}\label{eq:Int-Hamiltonian-Sch2}
    \hat{H}_{I}(T) &=& \hbar\,g(\partial\tau/\partial T)\Big(\phi_{\nu}\big[t(T),z(T)\big]\,\hat{a}_{\nu}+h.c.\Big)\,\big(e^{-i\,\omega_{T}T}\,\hat{\sigma}+h.c.\big)~;\\
    \mathcal{P}^{ex}_{\nu} (\omega_{T}) &=& g^2\bigg|\int_{T_{-}}^{T_{+}} dT\,(\partial\tau/\partial T)\,\phi^{\star}_{\nu}\big[t(T),z(T)\big]\,e^{i\,\omega_{T}T}\bigg|^2~.
    \label{eq:TP-gen-Sch2}
\end{eqnarray}
\end{subequations}
In the above scenario, we considered the atomic angular frequency to be $\omega_{T}$. In our subsequent analysis, we will see that the integration limits in the above expression $T_{-}$ and $T_{+}$ are finite as the Kruskal region is bounded by the central singularity and the horizon. It is to be noted that near the event horizon $r=r_{H}$ the factor $\partial\tau/\partial T\simeq 1$, and with this assumption $T_{\pm}$ will take certain values. We would like to mention that the above changes are expected to incorporate the essence of the background and the trajectories.

It is to be noted that for photons the interaction strength $g$ is time-independent, see \cite{Svidzinsky:2018jkp}, and thus we will treat it as a constant parameter in the estimation of the transition probability. In our subsequent analysis, we shall utilize the above expressions to estimate the transition probability for atoms following different trajectories and in different background spacetimes. Furthermore, we would like to mention that for the sake of brevity, we shall call all the different atomic angular frequencies $\omega$ in our subsequent analysis, as their differences in different scenarios are apparent. These $\omega$ will correspond to different $\omega_{\eta}$, $\omega_{T}$ etc. in appropriate scenarios. It is to be noted that when the atom is static in the Minkowski spacetime, this $\omega$ corresponds to the Minkowski frequency.

\section{Atomic excitations in Minkowski-light cone regions}\label{sec:Detector-response}
In this section, we examine two configurations involving an atom and a mirror in Minkowski spacetime, each described using different coordinate systems and notions of proper times.
Here, the mirror signifies a boundary in the spacetime from which the field modes get reflected, so that the whole mode function vanishes at the boundary. Mathematically, it is realized through specific boundary conditions on the quantum fields. For instance, with a Dirichlet boundary condition, the field modes vanish on the boundary, mimicking the behaviour of an ideal mirror. It is to be noted that these boundaries are crucial in cavity quantum electrodynamics (QED)~\cite{10.1093/acprof:oso/9780198509141.001.0001}, where confined field modes lead to enhanced light-matter interactions. They appear in theoretical models involving entanglement dynamics and radiative properties of atoms near surfaces~\cite{Messina_2007, PhysRevA.76.062114, PhysRevLett.91.243004, PhysRevA.76.032107}. They are also imperative in describing phenomena such as the dynamical Casimir effect~\cite{Hawking:1979ig, Moore:1970tmc}, providing insight into particle creation in the presence of time-dependent boundary conditions~\cite{Davies:1977yv, Davies:1976hi, Hotta:2015yla, Good:2016atu, Good:2018ell, Fernandez-Silvestre:2021ghq, Myrzakul:2021bgj, PhysRevD.94.065010, Zhou:2013hx}.
The presence of such a boundary breaks the Poincaré symmetry of spacetime and modifies the field's spectral density, thereby influencing the response of Unruh-DeWitt detectors~\cite{PhysRevD.14.870, PhysRevD.104.124001, Takagi:1986kn}. In summary, reflecting boundaries are important for both conceptual and practical purposes involving the study of vacuum fluctuations, quantum communication, and quantum thermodynamic cycles~\cite{Ng:2018drz, PhysRevD.94.064027, PhysRevD.93.024019, Arias:2015moa}.

In our present work, although both the atom and the mirror remain fixed in space within their respective coordinate systems, their time evolutions are governed by different notions of proper time—one associated with Minkowski time and the other with light cone time. We consider two distinct scenarios: In Case~(i), the atom is situated in the future light cone (FLC) region at $\zeta = 0$, with its proper time associated with the light cone time coordinate $\eta$. The mirror, on the other hand, is positioned at $z = z_0$ in Minkowski coordinates, with the fields represented in terms of the Minkowski time $t$ (see Fig.~\ref{fig:SD-Kasner}). In Case~(ii), the atom is placed at $z = z_0$ and evolves with proper time associated with Minkowski time $t$, while the mirror is located in the FLC region at $\zeta = 0$, with the fields represented in terms of the light cone time $\eta$ (see Fig.~\ref{fig:SD-Kasner}). We will also consider the $(1+1)$ and $(3+1)$ dimensional spacetimes separately to investigate and compare the above two scenarios as the spacetime dimensions change.

\subsection{$(1+1)$ dimensional scenario}\label{subsec:Mink-1p1}

Here, we calculate the transition probabilities of the atom from its ground state to its excited state for the above-mentioned Case (i) and Case (ii). In the calculation, whenever necessary, the relations Eq. \eqref{eq:CT-F} will be used.

\subsubsection{Case (i): Static mirror at $z=z_{0}$}\label{subsubsec:st-mirror-1p1}

In this scenario, the mirror is static in the Minkowski spacetime and the atom is in the future light cone region and we will study the atomic transition probability with respect to the conformal time $\eta$. In this regard, we consider the expression of Eq. \eqref{eq:TP-gen-kasner1} for the evaluation of the transition probability. From Eq. \eqref{eq:TP-gen-kasner1} it is evident that to investigate the transition probability one needs to find the expression of the scalar field mode in the Mirror's frame, i.e. in Minkowski spacetime. We shall express a single photon mode in $(1+1)$ dimensions in the presence of a reflecting mirror. In this regard, we consider the approaches as considered in \cite{book:Birrell, Svidzinsky:2018jkp}, and obtain
\begin{eqnarray}\label{eq:FM-1p1-AinF}
    \phi_{\nu} &=& \frac{1}{\sqrt{4\pi\,\nu}}\Big[e^{-i\,\nu\,t-i\,k(z-z_{0})}-e^{-i\,\nu\,t+i\,k(z-z_{0})}\Big]\nonumber\\
~&=& -\frac{2\,i\,\sin{\{\nu\,(z-z_{0})/c\}}}{\sqrt{4\pi\,\nu}}\,e^{-i\,\nu\,t}~.
\end{eqnarray}
As we are interested in massless field modes, the wave vector and the angular frequency are related as $\nu=|k|\,c$. This has been utilized in the last equality of Eq. \eqref{eq:FM-1p1-AinF}. The atom is in $FLC$, situated at $\zeta=0$ (i.e. at $z=0$). From Eq. \eqref{eq:CT-F} we observe that the time relevant for the atom is now $\eta$. More precisely the positive frequency modes with respect to the time $\eta$ corresponds to a conformal vacuum, a state devoid of any particles as seen by an observer with time $\eta$. 
Here the relation among the times along $\zeta=0$ is 
$t=\frac{c}{a} e^{\frac{a\eta}{c}}$  (see, Eq. \eqref{eq:CT-F}).
Then in the presence of a mirror, the transition probability of the atom due to the emission of the Minkowski photon will become 
\begin{eqnarray}\label{eq:TP-1p1-AinF-1}
    \mathcal{P}^{ex}_{\nu}(\omega) &=& \frac{4\,g^2}{4\pi\,\nu}\,\bigg|\int_{-\infty}^{\infty}d\eta\,e^{a\eta/c}\,e^{i\,\nu\,t}\,\sin{(\nu\,z_{0}/c)}\,e^{i\,\omega\,\eta}\bigg|^2\nonumber\\
    ~&=& \frac{2\,g^2\,c\,\omega\,\sin^2{(\nu\,z_{0}/c)}}{a\,\nu^{3}}\,\frac{1}{e^{2\pi\,c\,\omega/a}-1}~.
\end{eqnarray}

Now, we would like to mention that one can try to evaluate the transition probability due to the effect of all the field modes by integrating over $\nu$, see Appendix \ref{Appn:TP-mode-indp} regarding this. In particular, in the current scenario, by integrating over all the modes, we get
\begin{eqnarray}\label{eq:TP-1p1-AinF-2}
    \mathcal{P}^{ex}(\omega) &=& \int_{0}^{\infty} \mathbb{P}^{ex}_{\nu}(\omega)\,d\nu~,
\end{eqnarray}
where
\begin{eqnarray}\label{eq:TP-1p1-AinF-3}
    \mathbb{P}^{ex}_{\nu} (\omega) &=& \frac{4\,g^2\,\omega}{a}\,\frac{1}{e^{2\pi\,c\,\omega/a}-1}\,\frac{\sin^2{(\nu\,z_{0}/c)}}{\nu^{3}}~.
\end{eqnarray}
It should also be noted that the coupling constant $g$ is independent of $\tau$ (see \cite{Svidzinsky:2018jkp}), and we consider them to be independent of the field frequency $\nu$ as well. The quantity $\mathbb{P}^{ex}_{\nu} (\omega)$ is useful as the expression of this quantity can be compared between different scenarios, which will essentially enable one to compare the transition probability for a fixed mode frequency reminiscent of complete probability $\mathcal{P} ^{ex}(\omega)$. 
In particular, from the expression \eqref{eq:TP-1p1-AinF-3} we observe that, if $g$ is independent of $\nu$, the probability $\mathbb{P}^{ex}_{\nu} (\omega)$ is infrared divergent. This is evident as one takes the limits $\nu\to 0$ in expression \eqref{eq:TP-1p1-AinF-3}. At the same time, the probability $\mathbb{P}^{ex}_{\nu} (\omega)$ is not ultraviolet divergent, which can be checked by taking the limit $\nu\to \infty$. In particular, in the limit $\nu\to \infty$ the probability $\mathbb{P}^{ex}_{\nu} (\omega)\to 0$, which will also be evident if we evaluate the field-independent transition probability, see Eq. \eqref{Appeq:TP-1p1-AinF-4} of Appendix \ref{Appn:FieldIndp-Kasner}.

We would like to point out that the transition probability of Eq. \eqref{eq:TP-1p1-AinF-3} satisfies the KMS detailed balance condition, as can be checked by considering the change of variables $\omega\to -\omega$. The reason behind this is imprinted in the modes of \eqref{eq:FM-1p1-AinF} that undergo a coordinate transformation in \eqref{eq:TP-1p1-AinF-1}, analogous to the Rindler coordinate transformation. In our current scenario, the entire field mode is represented in either the Minkowski or light-cone coordinates, facilitating the appearance of a thermal transition probability. Thus, our analysis is analogous to evaluating Bogoliubov transformation coefficients between two modes, both of which are outgoing (or ingoing), but one in Minkowski and the other in the light cone region. On the contrary, in \cite{Obadia:2002ch}, an observer corresponding to the outgoing modes sees the ingoing modes, both in the Minkowski frame. In these works, only a part of the whole mode contains the signature of the mirror's accelerated motion, which leads to non-thermal spectra. We believe this crucial distinction between these works and ours leads to the differences in the respective transition probabilities. It is to be noted that non-thermal spectra are also encountered in the context of the Hawking effect in extremal black hole spacetimes (see \cite{Good:2020nmz, Barman:2018ina, Ghosh:2021ijv}), where the non-thermal outcomes are always attributed to the inverse relation between the ingoing and outgoing null coordinates in the Bogoliubov transformation.

\subsubsection{Case (ii): Static atom at $z=z_{0}$}\label{subsubsec:st-atom-1p1}

Here, we consider a static atom at $z=z_{0}$ which uses Minkowski coordinate time $t$, and the mirror is at $\zeta=0$ (i.e. at $z=0$) in the region of $FLC$. So, the photons are produced in a mirror frame, which is the light cone region here. Following the procedure presented in \cite{Svidzinsky:2018jkp}, we shall decompose the field modes in terms of the coordinates $(\eta, \zeta)$, i.e., the coordinates specific to the region $FLC$. Then, in $(1+1)$ dimensions, we have the scalar field mode solution as
\begin{eqnarray}\label{eq:FM-1p1-MinF}
    \phi_{\nu}(\eta,\zeta) &=& \frac{1}{\sqrt{4\pi\,\nu}}\,\Big[e^{-i\,\nu\,\eta+i\,k\,\zeta}-e^{-i\,\nu\,\eta-i\,k\,\zeta}\Big]~.
\end{eqnarray}
One should note that with the coordinates specific to the region $PLC$, also the above mode function remains the same, with $(\eta, \zeta)$ replaced by $(\bar{\eta}, \bar{\zeta})$. The mirror is fixed at $z=0$, and it corresponds to $\zeta=0$. Moreover, when $\zeta=0$, the above field mode vanishes, satisfying the required boundary condition. With the coordinate transformation from Eqs. \eqref{eq:CT-F} we have the relations
\begin{eqnarray}\label{eq:coord-rel-F}
    \zeta = \frac{c^2}{a} \, \tanh^{-1}\Big(\frac{z}{c\,t}\Big)~&,& ~~\eta = \frac{c}{2\,a} \log\Big[\frac{a^2}{c^4}(c^2t^2-z^2)\Big]~.
\end{eqnarray}
Then one can write down the field mode in terms of the Minkowski coordinates as
\begin{eqnarray}\label{eq:FM-1p1-MinF-2}
    \phi_{\nu}(t,z) &=& \frac{1}{\sqrt{4\pi\,\nu}}\,\bigg(\exp{\Big[-\frac{i\,\nu\,c}{a}\,\log{\Big\{\frac{a}{c^2}(c\,t-z)\Big\}}\Big]}\,\Theta(c\,t-z)-\exp{\Big[-\frac{i\,\nu\,c}{a}\,\log{\Big\{\frac{a}{c^2}(c\,t+z)\Big\}}\Big]}\,\Theta(c\,t+z)\bigg)~.\nonumber\\
\end{eqnarray}
In the above, the Lorentz-Heaviside $\Theta$-function has been introduced. This is because the atom is using Minkowski time, which covers $t\in (-\infty,\,\infty)$ and therefore it will interact with the photon mode, existing both in $FLC$ and $PLC$ regions. Then the transition probability, with the help of Eq. \eqref{eq:TP-gen-kasner2}, for an atom static at $z=z_{0}$ will be
\begin{eqnarray}\label{eq:TP-1p1-MinF-1}
    \mathcal{P}^{ex}_{\nu} (\omega) &=& g^2 \bigg|\int_{-\infty}^{\infty} dt~\phi^{\star}_{\nu}[t,z_{0}]~e^{i\,\omega\,t}\bigg|^2~\nonumber\\
    ~&=& \frac{2\,g^2\,c\,\sin^2{(\omega\,z_{0}/c)}}{\omega^2\,a}\, \frac{1}{e^{2\pi\,c\,\nu/a}-1}~.
\end{eqnarray}
One can follow the detailed steps provided in Appendix \ref{Appn:Pnu-stAtm-1p1} to obtain the last expression.

Now we would like to obtain the mode-independent transition probability (see Appendix \ref{Appn:TP-mode-indp}) by integrating over the field mode degrees of freedom, i.e., over the wave vector $k$. In particular, we obtain this expression as $\mathcal{P} ^{ex}(\omega) =  \int_{0}^{\infty}\,\mathbb{P}^{ex}_{\nu} (\omega)\,d\nu\,,$ where
\begin{eqnarray}\label{eq:TP-1p1-MinF-4}
    \mathbb{P}^{ex}_{\nu} (\omega) &=& \frac{4\,g^2\,\sin^2{(\omega\,z_{0}/c)}}{\omega^2\,a}\, \frac{1}{e^{2\pi\,c\,\nu/a}-1}~.
\end{eqnarray}
Like earlier, here also $\mathbb{P}^{ex}_{\nu} (\omega)$ signifies the transition probability for a fixed mode frequency, reminiscent of a mode-independent probability $\mathcal{P} ^{ex}(\omega)$. 
We consider $g$ to be time independent and also independent of the field frequency $\nu$. With these conditions, as one takes $\nu\to 0$, the above expression in Eq. \eqref{eq:TP-1p1-MinF-4}, like Eq. \eqref{eq:TP-1p1-AinF-3}, can be seen to be infrared divergent in terms of the field mode frequency. At the same time, in the ultraviolet limit $\nu\to \infty$, the transition probability $\mathbb{P}^{ex}_{\nu} (\omega)\to 0$, i.e., the above transition probability does not contain any UV divergence.

\subsection{$(3+1)$ dimensional scenario}\label{subsec:Mink-3p1}
We now do the same analysis in ($3+1$)-spacetime dimensions. This is a much more realistic situation than the previous ($1+1$)-dimensional analysis, as the mirror must be an extended object, and in reality, we are experiencing ($3+1$) dimensions. In $(3+1)$ dimensions, the coordinate transformations to regions $FLC$ and $PLC$ remain the same, like Eqs. \eqref{eq:CT-F} and \eqref{eq:CT-P}, with the spatial coordinates $x$ and $y$ now unchanged. Here also, we investigate both situations.

\subsubsection{Case (i): Static mirror at $z=z_{0}$}\label{subsubsec:st-mirror-3p1}

For a static mirror at $z=z_{0}$, i.e., on the $x-y$ plane, the field modes should be evaluated in terms of the Minkowski coordinates. This field mode solution is straightforward to estimate \cite{book:Birrell}, and is expressed as
\begin{eqnarray}\label{eq:FM-3p1-AinF}
    \phi_{\nu,\vec{k}_{\perp}} &=& \frac{1}{\sqrt{(2\pi)^3\,2\,\nu}}e^{-i\nu\,t+i\,\vec{k}_{\perp}.\vec{x}_{\perp}}\Big[e^{-i\,k_{z}(z-z_{0})}-e^{i\,k_{z}(z-z_{0})}\Big]~\nonumber\\
    ~&=& -\frac{2\,i\,\sin{\{k_{z}(z-z_{0})\}}}{\sqrt{(2\pi)^3\,2\,\nu}}e^{-i\nu\,t+i\,\vec{k}_{\perp}.\vec{x}_{\perp}}~.
\end{eqnarray}
In Eq. \eqref{eq:FM-3p1-AinF}, one can observe that at $z=z_{0}$, i.e., at the position of the mirror, the field mode vanishes. Here also, we consider the atom to be in the future light cone region $FLC$ at $\zeta=0$.
With this consideration, we use the expression of the above mode solution in Eq. \eqref{eq:TP-gen-kasner1} and get the transition probability as 
\begin{eqnarray}\label{eq:TP-3p1-AinF-1}
    \mathcal{P}^{ex}_{\nu} (\omega) &=& \frac{4\,g^2\,\sin^2{(k_{z}z_{0})}}{(2\pi)^3\,2\,\nu}\,\bigg|\int_{-\infty}^{\infty}d\eta\,e^{a\eta/c}\,e^{i\,\nu\,t+i\,\omega\,\eta}\bigg|^2\nonumber\\
    ~&=& \frac{g^2\,c\,\omega\,\sin^2{(k_{z}z_{0})}}{2\pi^2\,\nu^{3} \,a}\,\frac{1}{e^{2\pi\,c\,\omega/a}-1}~.
\end{eqnarray}

Now we integrate the above expression over the field wave vectors to obtain a field-independent result, like the prescription provided in Appendix \ref{Appn:TP-mode-indp}. Then one finds $\mathcal{P} ^{ex}(\omega) = \int_{0}^{\infty} \nu^2 d\nu \int_{0}^{\pi} \sin\theta\, d\theta \int_{0}^{2\pi} d\phi~  \mathcal{P}^{ex}_{\nu} (\omega)
= \int_{0}^{\infty} \mathbb{P}^{ex}_{\nu} (\omega) \,d\nu\,$. With the identification $k_{z}=\nu\,\cos{\theta}$, we can obtain
\begin{eqnarray}\label{eq:TP-3p1-AinF-3}
    \mathbb{P}^{ex}_{\nu} (\omega) &=& \frac{g^2\,\omega}{\pi\,a\,c^2}\,\frac{1}{e^{2\pi\,c\,\omega/a}-1}\,\bigg[1 -\frac{c}{2\,z_{0}\,\nu}\sin{\Big(\frac{2\,z_{0}\,\nu}{c}\Big)}\bigg]\,\frac{1}{\nu} ~.
\end{eqnarray}
From this expression, one can observe that as $\nu\to \infty$ the probability $\mathbb{P}^{ex}_{\nu} (\omega)\to 0$. Therefore, this probability is not ultraviolet (UV) divergent, just like the expressions of $\mathbb{P}^{ex}_{\nu} (\omega)$ from Eqs. \eqref{eq:TP-1p1-AinF-3} and \eqref{eq:TP-1p1-MinF-4}. At the same time, in the infrared limit $\nu\to 0$, the transition probability $\mathbb{P}^{ex}_{\nu} (\omega)$ diverges out. We would also like to mention that for large $a$, especially when $c\,\omega/a\ll 1$, the above expression for transition probability becomes independent of $a$. In this scenario, the above transition probability can be approximated to be $\mathbb{P}^{ex}_{\nu} (\omega)= g^2\nu[1-\sin{(z_{\nu})}/z_{\nu}]/\{2c(\pi\,c\,\nu)^2\}$, where $z_{\nu}=2\,z_{0}\,\nu/c$.

Here also the transition probability \eqref{eq:TP-3p1-AinF-3} satisfies the KMS detailed balance condition, much like the $(1+1)$ dimensional scenario. The reason is that the relevant mode of Eq. \eqref{eq:FM-3p1-AinF}, in its entirety, gets transformed to light-cone coordinates, which is analogous to the Rindler coordinate transformation. In contrast, in \cite{Obadia:2002ch, Good:2020nmz}, a part of the whole mode contains the signature of the accelerated motion leading to non-thermal spectra.

\subsubsection{Case (ii): Static atom at $z=z_{0}$}\label{subsubsec:st-atom-3p1}

Here we consider the situation where the atom is kept static at $z=z_{0}$, and the mirror is at $\zeta=0$ spanning from $PLC$ to $FLC$. In $(3+1)$ dimensional light cone region of Minkowski spacetime, we have the field mode solution, see Appendix \ref{Appn:Mode-3p1-F}, as
\begin{eqnarray}\label{eq:FM-3p1-AinF-gen}
    \phi_{\nu,\vec{k}_{\perp}}(\eta, \zeta) &=& \sqrt{\frac{c\,\cosech(\pi\,c\,\nu/a)}{16\pi^2\,a}}\,\mathrm{J}_{-\frac{i\,c\,\nu}{a}}\bigg(\frac{c^2k_{\perp}\,e^{a\,\eta/c}}{a}\bigg) \,e^{\frac{i\,\nu\,\zeta}{c}+i\,\Vec{k}_{\perp}.\Vec{x}_{\perp}}~,
\end{eqnarray}
where $\mathrm{J}_{n}(x)$ signifies the Bessel function of first kind of order $n$, and $\vec{x}_{\perp}=(x,y)$ is the spatial distance in the transverse direction. For analytical calculation, we will consider a particular choice of $a$. For large $a$, one can expand the Bessel function in series as, see \cite{NIST:DLMF},
\begin{eqnarray}\label{eq:BesselJ-series}
    \mathrm{J}_{-\frac{i\,c\,\nu}{a}}\bigg(\frac{c^2k_{\perp}\,e^{a\,\eta/c}}{a}\bigg) \simeq -\frac{i\,a}{c\,\nu} \Big(\frac{c^2\,k_{\perp}}{2\,a}\Big)^{-i\,\nu\,c/a}\,\frac{e^{-i\,\nu\,\eta}}{\Gamma(i\,\nu\,c/a)}~.
\end{eqnarray}
From Eq. \eqref{eq:BesselJ-series} one can observe that the field mode of \eqref{eq:FM-3p1-AinF-gen} has an outgoing characteristic, as in large $a$ limit the field mode behaves like $\phi_{\nu,\vec{k}_{\perp}}(\eta, \zeta)\sim e^{-i\,\nu\,\eta+i\,\nu\,\zeta/c}$. We follow the prescription as provided in \cite{Svidzinsky:2018jkp}, according to which the field mode function $\phi_{\nu,\vec{k}_{\perp}}(t,z)$ in the atom's frame is obtained from $\phi_{\nu,\vec{k}_{\perp}}(t,z)=\{\phi_{\nu,\vec{k}_{\perp}}(\eta, \zeta)-\phi_{\nu,\vec{k}_{\perp}}(-\eta, \zeta)\}$ utilizing the coordinate transformation from $(\eta, \zeta)$ to $(t,z)$ of Eq.  \eqref{eq:coord-rel-F}. In that scenario, $\phi_{\nu,\vec{k}_{\perp}}(t,z)$ will contain both the outgoing and ingoing mode functions. It should also be noted that the outgoing part of this field mode will interact with the atom in the FLC region, i.e., when $c\,t>z$. At the same time, the ingoing part of the mode function will interact with the atom only in the PLC region, i.e., when $c\,t>-z$. Therefore, these outgoing and ingoing parts of the mode functions should be accompanied by the Heaviside step functions $\Theta (ct-z)$ and $\Theta (ct+z)$, respectively (this idea will be used in the latter part of the investigation as well). Then, in the large $a$ limit, the field mode in the atom's frame can be expressed as
\begin{eqnarray}\label{eq:FM-3p1-MinF}
\phi_{\nu,\vec{k}_{\perp}}(t,z) &\simeq& \sqrt{\frac{c\,\cosech(\pi\,c\,\nu/a)}{16\pi^2\,a}}\,\Big(-\frac{i\,a}{c\,\nu}\Big) \Big(\frac{c^2\,k_{\perp}}{2\,a}\Big)^{-i\,\nu\,c/a}\,\frac{1}{\Gamma(i\,\nu\,c/a)}\nonumber\\
~&\times& \bigg(\exp{\Big[-\frac{i\,\nu\,c}{a}\,\log{\Big\{\frac{a}{c^2}(c\,t-z)\Big\}}\Big]}\,\Theta(c\,t-z)\nonumber\\
    ~&& -\exp{\Big[-\frac{i\,\nu\,c}{a}\,\log{\Big\{\frac{a}{c^2}(c\,t+z)\Big\}}\Big]}\,\Theta(c\,t+z)\bigg)~.\
\end{eqnarray}
The transition probability, with the help of Eq. \eqref{eq:TP-gen-kasner2}, for the atom static at $z=z_{0}$ will be
\begin{eqnarray}\label{eq:TP-3p1-MinF-1}
    \mathcal{P}^{ex}_{\nu} (\omega) &=& \frac{g^2\,c\,\sin^2{(\omega\,z_{0}/c)}}{2\,\pi^2\,\omega^2\,a}\, \frac{1}{e^{2\pi\,c\,\nu/a}-1}~.
\end{eqnarray}
To follow the explicit steps leading to the last expression, one can check Appendix \ref{Appn:Pnu-stAtm-3p1}.

From the above expression, we can obtain the transition probability for all photon modes by integrating over the field wave vectors. In this regard, we follow the prescription provided in Appendix \ref{Appn:TP-mode-indp}. In particular, the field-independent transition probability corresponding to the above expression is obtained as $\mathcal{P} ^{ex}(\omega) =  \int_{0}^{\infty}\mathbb{P}^{ex}_{\nu} (\omega)\,d\nu\,,$ where we have
\begin{eqnarray}\label{eq:TP-3p1-MinF-4}
    \mathbb{P}^{ex}_{\nu} (\omega) &=& \frac{2\,g^2\,\sin^2{(\omega\,z_{0}/c)}}{\pi\,\omega^2\,a\,c^2}\,\frac{\nu^2}{e^{2\pi\,c\,\nu/a}-1}~.
\end{eqnarray}
One can observe that $\mathbb{P}^{ex}_{\nu} (\omega)$ is finite in the limit of $\nu\to 0$ and $\nu\to \infty$ for $g$ being independent of $\nu$. In particular, when $\nu\to 0$ the behaviour of the above probability is determined by the numerator $\nu^2$, which will result in $\mathbb{P}^{ex}_{\nu} (\omega)$ becoming zero. At the same time, when $\nu\to \infty$ the behaviour of $\mathbb{P}^{ex}_{\nu} (\omega)$ is determined by the denominator, which will again make the quantity zero. These results are in contrast with the one obtained in Eq. \eqref{eq:TP-3p1-AinF-3} corresponding to a static mirror at $z=z_{0}$ in $(3+1)$ dimensions. We would also like to mention that for large $a$, more precisely when $c\,\nu/a\ll 1$, the above expression of transition probability simplifies to $\mathbb{P}^{ex}_{\nu}(\omega)= g^2\nu^2\sin^2{[\omega\,z_{0}/c]}/\{c\,(\pi\,\omega\,c)^2\}$.

In this section, we found that due to relative time translation between a two-level atom and a mirror, the photon in one frame enables the atomic transition. It may be noted that classically, the relative time translations between the Minkowski frame and light cone frame are equivalent to those of the respective frames in reverse. However, whether this symmetry is still valid in the quantum regime needed investigation. Our analysis in ($1+1$)-dimensions showed that the transition probabilities (\ref{eq:TP-1p1-AinF-3}) and (\ref{eq:TP-1p1-MinF-4}) are equal when the photon and atom frequencies are the same. Moreover, only the thermal factors for both situations show symmetry under $\nu \leftrightarrow \omega$. However, the situation in ($3+1$) dimensions is different. Although $\nu \leftrightarrow \omega$ symmetry exists in thermal factors, the whole transition probabilities are not the same when one takes $\nu=\omega$ (see, Eqs. \eqref{eq:TP-3p1-MinF-4} and \eqref{eq:TP-3p1-AinF-3}). 
It is to be noted that when we take the large $a$ limit, the transition probabilities become independent of $a$ for both scenarios. However, even at this limit, they do not have a similar functional form.
In $(3+1)$ dimensions, one can identify the origin of this absence of symmetry in the presence of $k_{z}$ in Eq. \eqref{eq:TP-3p1-AinF-1}. The presence of the specific field wave vector $k_{z}$ again originates from the directional dependence of quantum transitions on the location of the mirror. Moreover, the transition probabilities for all photon modes do not have similarities when comparing between two situations, Case (i) and Case (ii). Therefore, it seems that the classical symmetry in the relative time translations is not reflected in the atom's transition probabilities when comparing two Killing frames in $(3+1)$ dimensions. 
We believe that the robustness of this aforesaid equivalence is valid only in $(1+1)$ dimensions. The reason is that, in $(1+1)$ dimensions, the field modes in both Minkowski and light cone regions are plane wave solutions. This structural similarity between the field modes results in the equivalence in atomic transition probabilities between different scenarios in $(1+1)$ dimensions. At the same time, in $(3+1)$ dimensions, the Minkowski and the light cone field modes are not structurally analogous, as the line element in the light cone region is not conformally flat.

\section{Atomic excitations for black hole spacetime}\label{sec:Sch-transition}

In this section, we focus on atomic transitions in a Schwarzschild black hole spacetime. The inside region of the black hole has a close resemblance to the Minkowski light cone region when one describes the inside region through KS and Schwarzschild coordinates. Therefore, we consider two particular scenarios by keeping either the atom or the mirror fixed at a spatial point. Two timelike coordinates are defined within the BH -- one is KS time, and the other is the radial coordinate. Correspondingly, two time frames can be associated with the mirror and the atom, respectively.  Two scenarios respectively correspond to -- 
Case $(i)$, the mirror frame contains field modes parametrized by the Kruskal--Szekeres (KS) coordinates $(X, T)$. The mirror remains fixed at $X = X_0$. The atom, in contrast, is using Schwarzschild coordinates $(t_{S}, r_{\star})$ inside the black hole. It evolves along a trajectory with constant $t_{S} = 0$, which corresponds to $X = 0$ in KS coordinates. Its proper time is associated with the timelike coordinate $r_{\star}/c$. Case~$(ii)$, the configuration is reversed. The atom is using KS coordinates and remains fixed at $X = X_0$, with its proper time related to the KS time $T$. The mirror, on the other hand, is described by Schwarzschild coordinates inside the black hole. It evolves along a trajectory with $t_{S} = 0$, and its proper time is associated with the timelike coordinate $r_{\star}/c$.
 Our motivation behind considering this particular system is twofold. These two situations are classically identical with respect to the atomic frame. Therefore, like earlier, we want to investigate whether the aforesaid symmetry still holds in the atomic transition. Secondly, the future KS region has an apparent analogy with FLC. Therefore, like RRW and outside BH region correspondence, we want to check whether the transitions inside BH have any resemblance to those in the light cone regions. However, from Figs. \ref{fig:SD-Kasner} and \ref{fig:SD-Kruskal}, it can be seen that the interior of a Schwarzschild black hole is not entirely analogous to the light cone region due to the presence of a singularity at $r=0$. Therefore, we expect some modifications in the transition probability compared to the light cone region. We would also like to mention that we will be dealing with a $(1+1)$ dimensional Schwarzschild black hole spacetime, as in $(3+1)$ dimensions, the exact functional form of the field mode solution is not known. Moreover, from \cite{Barman:2017fzh}, one can see that though these modes in near-horizon and asymptotic regions become plane wave-like, their behaviour in the $r\to 0$ limit is not well defined.

\subsection{Case (i): The Mirror is at $X=X_{0}$}
In this scenario, the modes interacting with the mirror are expressed in terms of the Kruskal--Szekeres (KS) coordinates, and the mirror is positioned at a constant spatial location $X = X_0$, with its evolution governed by the KS time coordinate $T$. In contrast, the atom is represented in Schwarzschild coordinates inside the black hole, traversing a path along which the spacelike coordinate $t_{S}$ remains fixed at zero. This trajectory corresponds to $X = 0$ in the KS framework, and the atom's proper time progresses according to the timelike coordinate $r_{\star}/c$.
So, the atom will transit from the ground state to the excited state when encountering the produced photon in the KS frame. Therefore, we describe the photon modes in terms of the KS coordinates, with the vacuum given by the conformal vacuum relating to the positive frequency mode $e^{-i\,\nu\,T}$. In particular, this single-photon mode solution, see \cite{book:Birrell, Svidzinsky:2018jkp}, is given by
\begin{eqnarray}\label{eq:FM-Sch-Kruskal}
    \phi_{\nu} &=& \frac{1}{\sqrt{4\,\pi\,\nu}}\,e^{-i\,\nu\,T}\Big[e^{-i\,|k|(X-X_{0})}-e^{i\,|k|(X-X_{0})}\Big]~.
\end{eqnarray}
Utilizing this mode solution in the expression of the transition probability of Eq. \eqref{eq:TP-gen-Sch1}, we obtain
\begin{eqnarray}\label{eq:TP-Sch-MatX0-3}
    \mathcal{P}^{ex}_{\nu} (\omega) 
    ~&=& \frac{2\,g^2\,\omega\,\sin^{2}(\nu\,X_{0}/c)}{\nu^{3}\,\kappa_{H}\,c}\,\frac{1}{e^{2\,\pi\,\omega/(c\,\kappa_{H})}-1}\,\left|1-\frac{\Gamma\Big(\frac{i\,\omega}{c\,\kappa_{H}}+1,-\frac{i\,\nu\,\chi^{f}}{c\,\kappa_{H}}\Big)}{\Gamma\Big(\frac{i\,\omega}{c\,\kappa_{H}}+1\Big)}\right|^2~.
\end{eqnarray}
One can follow the exact steps leading to this last expression of the excitation probability in Appendix \ref{Appn:Pnu-stMirr-BH}. We would also like to mention that in this last expression $\chi^{f}=e^{-\kappa_{H}\,r_{\star}^{f}}$, where $r_{\star}^{f}$ is a finite value of $r_{\star}$ up to which the relevant integral over $r_{\star}$ is carried out. It is to be noted that ideally, the integration over $r_{\star}$ should be carried out till zero, which corresponds to the Schwarzschild inner singularity. However, one cannot obtain analytical results, as per our understanding, from this integration. At the same time, we are more interested in atomic transitions in the near-horizon region as we want to investigate the quantum equivalence between different scenarios. Thus, the limit $r_{\star}^{f}$ for the $r_{\star}$ integration becomes a valid approximation to provide the relevant results in the near-horizon region, which also happens to provide analytical expressions for the atomic excitation.

From this last expression \eqref{eq:TP-Sch-MatX0-3} one can obtain the expression of mode-independent transition probability by integrating over all field modes, see Appendix \ref{Appn:TP-mode-indp}. The expression of this mode-independent transition probability is $\mathcal{P} ^{ex}(\omega) =  \int_{0}^{\infty}\,\mathbb{P}^{ex}_{\nu} (\omega)\,d\nu\,,$ where $\mathbb{P}^{ex}_{\nu} (\omega)$ is given by
\begin{eqnarray}\label{eq:TP-Sch-MatX0-5}
    \mathbb{P}^{ex}_{\nu} (\omega) = \frac{4\,g^2\,\omega\,\sin^{2}(\nu\,X_{0}/c)}{\nu^{3}\,\kappa_{H}\,c^2}\,\frac{1}{e^{2\,\pi\,\omega/(c\,\kappa_{H})}-1}\,\left|1-\frac{\Gamma\Big(\frac{i\,\omega}{c\,\kappa_{H}}+1,-\frac{i\,\nu\,\chi^{f}}{c\,\kappa_{H}}\Big)}{\Gamma\Big(\frac{i\,\omega}{c\,\kappa_{H}}+1\Big)}\right|^2~.
\end{eqnarray}
If we consider $g$ to be independent of $\nu$, then the probability $\mathbb{P}^{ex}_{\nu} (\omega)$ in the limit of $\nu\to 0$ is $\mathbb{P}^{ex}_{\nu} (\omega)\sim \nu$. At the same time, in the limit $\nu\to \infty$, this probability becomes $\mathbb{P}^{ex}_{\nu} (\omega)\sim \mathcal{O}(1/\nu^3)$. Therefore, in this scenario, the transition probability $\mathbb{P}^{ex}_{\nu} (\omega)$ does not diverge in the infrared or ultraviolet limits of the field frequency if $g$ remains finite in the same limits.

\subsection{Case (ii): The atom is at $X=X_{0}$}
Here, we consider the atom to be static at $X= X_{0}$, and it follows the KS time $T$. The mirror is now at $t_{S}=0$ axis (i.e. at $X=0$) and using $r_\star/c$ as its coordinate time. So, the atom will encounter the photons produced in the frame described by interior Schwarzschild coordinates. We follow a procedure similar to the one proposed in \cite{Svidzinsky:2018jkp} and obtain the concerned photon mode as
\begin{eqnarray}\label{eq:FM-1p1-MinF-3}
    \phi_{\nu}(r_{\star},t_{S}) &=& \frac{1}{\sqrt{4\pi\,\nu}}\,\bigg[e^{i\,\nu\,u/c}\,\Theta(c\,T-X)-e^{i\,\nu\,v/c}\,\Theta(c\,T+X)\bigg]~\nonumber\\
    ~&=& \frac{1}{\sqrt{4\pi\,\nu}}\,\bigg[\big\{\kappa_{H}(c\,T-X)\big\}^{i\,\nu/(\kappa_{H}\,c)}\,\Theta(c\,T-X)-\big\{\kappa_{H}(c\,T+X)\big\}^{i\,\nu/(\kappa_{H}\,c)}\,\Theta(c\,T+X)\bigg]~.
\end{eqnarray}
In the last line of the previous expression, we have used the relation of Eq. \eqref{eq:Sch-Kruskal-EF}. The $\Theta$ functions have been introduced as the atom is using KS time, and so both future KS and past KS are accessible to the atom. Then the transition probability of the atom fixed at the KS position $X=X_{0}$, with the help of Eq. \eqref{eq:TP-gen-Sch2}, is given by
\begin{eqnarray}\label{eq:TP-Sch-AatX0-2}
    \mathcal{P}^{ex}_{\nu} (\omega) 
    &=& \frac{2\,g^2\,\sin^2{(\omega\,X_{0}/c)}}{\omega^2\,\kappa_{H}\,c}\,\frac{1}{e^{\frac{2\,\pi\,\nu}{c\,\kappa_{H}}}-1}\,\big|1-\Psi_{Sch}\big|^2~.
\end{eqnarray}
In the last expression $\Psi_{Sch}$ is given by
\begin{eqnarray}\label{eq:TP-Sch-AatX0-Psi}
    \Psi_{Sch} &\equiv& \Psi_{Sch}(\omega,\,\nu)\nonumber\\
    ~&=& \frac{1}{\Gamma\left(1+\frac{i\,\nu }{c\,\kappa_{H}}\right)}\Bigg[\Gamma \left(1+\frac{i\,\nu }{c\,\kappa_{H}},-\frac{i\,\omega (T_{0}-\kappa_{H}\,X_{0})}{c\,\kappa_{H}}\right)\nonumber\\
    ~&+& \frac{e^{-i\,\omega\,X_{0}/c}}{2\,i\,\sin{(\omega\,X_{0}/c)}}\,\bigg\{\Gamma \left(1+\frac{i\,\nu }{c\,\kappa_{H}},-\frac{i\,\omega (T_{0}+\kappa_{H}\,X_{0})}{c\,\kappa_{H}}\right)-\Gamma \left(1+\frac{i\,\nu }{c\,\kappa_{H}},-\frac{i\,\omega (T_{0}-\kappa_{H}\,X_{0})}{c\,\kappa_{H}}\right)\bigg\}\Bigg]~,
\end{eqnarray}
where $T_{0}=(e^{2\kappa_{H}\, r_{\star}^{f}} + \kappa_{H}^2\, X_{0}^2)^{1/2}$. The details to arrive at this expression are provided in Appendix \ref{Appn:Pnu-stAtm-BH}.\vspace{0.2cm}

Here also the mode-independent transition probability, according to Appendix \ref{Appn:TP-mode-indp}, can be expressed as $\mathcal{P} ^{ex}(\omega)=\int_{0}^{\infty}\,\mathbb{P}^{ex}_{\nu} (\omega)\,d\nu$, where $\mathbb{P}^{ex}_{\nu} (\omega)$ is now
\begin{eqnarray}\label{eq:TP-Sch-AatX0-3}
    \mathbb{P}^{ex}_{\nu} (\omega) &=& \frac{4\,g^2\,\sin^2{(\omega\,X_{0}/c)}}{\omega^2\,\kappa_{H}\,c^2}\,\frac{1}{e^{\frac{2\,\pi\,\nu}{c\,\kappa_{H}}}-1}\,\big|1-\Psi_{Sch}\big|^2~.
\end{eqnarray}
The above quantity in the limit of $\nu\to 0$ behaves as $\mathbb{P}^{ex}_{\nu} (\omega)\sim g^2/\nu$, which is divergent if $g$ is independent of the field frequency; i.e. it contains IR divergence, usually appears in two spacetime dimensions. However, $\mathbb{P}^{ex}_{\nu} (\omega)$ remains well behaved in the limit of $\nu\to \infty$ if $g$ is independent of the field frequency $\nu$.

Now we would like to compare Eqs. (\ref{eq:TP-Sch-MatX0-5}) and (\ref{eq:TP-Sch-AatX0-3}). Note that the thermal factors in both the expressions satisfy the symmetry under the transformation $\nu\to\omega$, whereas the full expressions are not the same when one chooses $\nu=\omega$. This contradicts the $(1+1)$-dimensional Minkowski-light cone analysis. Thus, the present analysis shows that though from the atom's frame both the cases are equivalent in terms of classical motion, the quantum phenomenon is not. 
One can notice that this disparate feature in the Schwarzschild transition probabilities arrives due to the presence of the terms $\Gamma\big[i\,\omega/(c\,\kappa_{H})+1,-i\,\nu\,\chi^{f}/(c\,\kappa_{H})\big]/ \Gamma\big[i\,\omega/(c\,\kappa_{H})+1\big]$ in \eqref{eq:TP-Sch-MatX0-5} and $\Psi_{Sch}$ in \eqref{eq:TP-Sch-AatX0-3}. Both of these terms will vanish if the interior of the Schwarzschild black hole is not bounded by the finite $r$ surface and rather if it extends to infinity, which is not possible, as even without the near-horizon approximation, the radial coordinate is bounded by $r=0$ inside the horizon.

The above observation also suggests that the present results are different from the $(1+1)$ dimensional Minkowski-light cone scenario. We would like to point out a specific difference between these two scenarios, which we believe leads to these different observations. For the atom confined to the light cone regions, the redshift factor is given by $\partial\tau/\partial\eta = e^{a\eta/c}$. Whereas, for the atom inside the Schwarzschild black hole and using the Schwarzschild coordinates, the redshift factor is $\partial\tau/\partial(r_{\star}/c) = (-1+r_{H}/r)^{1/2}$, which becomes similar to the light cone scenario only in the near-horizon region, i.e., $\partial\tau/\partial(r_{\star}/c) \simeq e^{\kappa_{H}\,r_{\star}}$ near the event horizon $r=r_{H}$. A similar situation is observed when the atom is using the Kruskal coordinates. One can notice that, though the coordinate transformations from Minkowski to light cone and Kruskal to Schwarzschild are analogous, the metric coefficients in the two spacetimes are not functionally similar, which leads to the differences in the redshift factors. Thus, for coordinate transformation to the light cone regions, the conformal time coordinate $\eta$ has range $(-\infty,\,\infty)$. At the same time, we are compelled to take the range of the tortoise coordinate $r_{\star}/c$, which mimics the conformal time inside the Schwarzschild event horizon, to be $(-\infty,\,r_{\star}^{f}]$. Here $r_{\star}^{f}<<0$ signifies a pivotal value of $r_{\star}$ near the event horizon that ensures the validity of the approximation $\partial\tau/\partial(r_{\star}/c) \simeq e^{\kappa_{H}\,r_{\star}}$. Therefore, the atomic transition probabilities corresponding to different backgrounds differ as we restrict ourselves to the near-horizon region in the BH background.

\section{Atomic de-excitations and excitation to de-excitation ratio}\label{sec:de-excitation}
In this section, we find the de-excitation probabilities of atoms corresponding to each previously discussed scenario. Subsequently, we also estimate the excitation to de-excitation ratio (EDR) for individual scenarios. It is observed in \cite{Kumawat:2024kul} that even if the individual atomic excitation probabilities corresponding to physically equivalent scenarios are not equivalent, the EDRs can indicate similarity, which in turn can provide crucial physical insights, as in a realistic scenario, it is more plausible to identify the EDRs corresponding to an atomic system than excitations or de-excitations. This motivated us to investigate the EDRs, also in our current work. In the following few subsections, we shall first estimate the de-excitation probabilities in the light cone and Kruskal regions, and then the EDRs corresponding to each scenario.

\subsection{Atomic De-Excitation in Minkowski-light cone regions}
In this part of the section, we investigate the atomic de-excitations in the light cone region of the Minkowski spacetime. In this regard, we first consider the $(1+1)$ dimensional scenario and then the $(3+1)$ dimensional scenario. In both dimensions, we shall also consider the situations with either the atom following the light cone time or the mirror following the light cone time.

In particular, with the help of the Hamiltonian defined in Eq. \eqref{eq:Int-Hamiltonian-kasner1} and the analysis from \cite{Kumawat:2024kul} one can obtain the formula of the de-excitation probability corresponding to an atom following the light cone time as
\begin{eqnarray}\label{eq:DE-gen-kasner1}
    \mathcal{P}^{de}_{\nu} (\omega_{\eta}) &=& g^2\bigg|\int_{-\infty}^{\infty} d\eta\,e^{a\,\eta/c}\,\phi^{\star}_{\nu}\big[t(\eta),z(\eta)\big]\,e^{-i\,\omega_{\eta}\,\eta}\bigg|^2~.
\end{eqnarray}
Similarly, for an atom following the Minkowski time and the mirror following the light cone time, we consider the appropriate Hamiltonian from Eq. \eqref{eq:Int-Hamiltonian-kasner2}. Then, with the help of Eq. \eqref{eq:TP-gen-kasner2} and the analysis from \cite{Kumawat:2024kul} we obtain the formula for the de-excitation probability as 
\begin{eqnarray}
    \mathcal{P}^{de}_{\nu} (\omega_{t}) &=& g^2\bigg|\int_{-\infty}^{\infty} dt\,\phi^{\star}_{\nu}\big[t,z\big]\,e^{-i\,\omega_{t}\,t}\bigg|^2~.
    \label{eq:DE-gen-kasner2}
\end{eqnarray}
In our following analysis, we shall utilize these two formulas from Eqs. \eqref{eq:DE-gen-kasner1} and \eqref{eq:DE-gen-kasner2} to obtain the de-excitation probabilities in $(1+1)$ and $(3+1)$ dimensions respectively.

\subsubsection{$(1+1)$ dimensional scenario}
\noindent \emph{Case (i): Static mirror at $z=z_{0}$ :-$~~$}
Here, we investigate the scenario in which an atom is in the light cone region at $\zeta=0$, and a mirror is static in Minkowski spacetime at $z=z_{0}$, i.e., the atom is following the light cone time and the field modes \eqref{eq:FM-1p1-AinF} are represented in terms of the Minkowski time. Combining this field mode solution with equation (\ref{eq:DE-gen-kasner1}) for the de-excitation probability we obtain
\begin{eqnarray}\label{eq:TP-DeEx-1p1-AinF-1}
    \mathcal{P}_{\nu}^{de}(\omega) &=& \frac{4\,g^2}{4\pi\,\nu}\,\bigg|\int_{-\infty}^{\infty}d\eta\,e^{a\eta/c}\,e^{i\,\nu\,t}\,\sin{(\nu\,z_{0}/c)}\,e^{-i\,\omega\,\eta}\bigg|^2\nonumber\\
    ~&=& \frac{2\,g^2\,c\,\omega\,\sin^2{(\nu\,z_{0}/c)}}{a\,\nu^{3}}\,\frac{1}{1-e^{-2\pi\,c\,\omega/a}}~.
\end{eqnarray}
One can compare this result with the $(1+1)$ dimensional result for excitation from Eq. \eqref{eq:TP-1p1-AinF-1}, and observe that by making $\omega\to -\omega$ in \eqref{eq:TP-1p1-AinF-1} one can obtain the de-excitation result. This is a crucial observation which may not hold true in other scenarios.\vspace{0.2cm}

\noindent \emph{Case (ii): Static atom at $z=z_{0}$ :-$~~$}
Here, we investigate the case where the atom is static in Minkowski spacetime at $z=z_0$, and the mirror is in the light cone region, situated at $\zeta=0$. Here, the atom is following the Minkowski time. At the same time, the field mode solution for this scenario, obtained in terms of the light cone coordinates and then represented in terms of the Minkowski coordinates, is provided in Eq. \eqref{eq:FM-1p1-MinF-2}. Utilizing this field mode solution with the de-excitation probability equation (\ref{eq:DE-gen-kasner2}), we obtain
\begin{eqnarray}\label{eq:TP-DeEx-1p1-MinF-1}
    \mathcal{P}_{\nu}^{de} (\omega) &=& g^2 \bigg|\int_{-\infty}^{\infty} dt~\phi^{\star}_{\nu}[t,z_{0}]~e^{-i\,\omega\,t}\bigg|^2~\nonumber\\
    ~&=&\frac{2\,g^2\,c\,\sin^2{(\omega\,z_{0}/c)}}{\omega^2\,a}\, \frac{1}{1-e^{-2\pi\,c\,\nu/a}}~.
\end{eqnarray}
One can observe the significant resemblance of this expression with $(1+1)$ dimensional excitation probability from Eq. \eqref{eqAppn:TP-1p1-MinF-2}. However, unlike the atom following the light cone time scenario, in this case, one cannot obtain the de-excitation probability by making $\omega\to -\omega$ in Eq. \eqref{eqAppn:TP-1p1-MinF-2}. This de-excitation probability cannot be obtained by even making  $\nu\to -\nu$ in the excitation probability expression. We would also like to mention that, similar to the excitation probabilities, here also, the de-excitation probabilities corresponding to the two different scenarios from Eqs. \eqref{eq:TP-DeEx-1p1-AinF-1} and \eqref{eq:TP-DeEx-1p1-MinF-1} become the same when atomic and field frequencies are equal, i.e., when $\omega=\nu$.

\subsubsection{$(3+1)$ dimensional scenario}

\noindent \emph{Case (i): Static mirror at $z=z_{0}$:-$~~$}
The situation where the mirror is static  $z=z_{0}$ on the $x-y$ plane and the atom is static in the future light cone region FLC $\zeta=0$ is examined here, i.e., the atom following the light cone time. In this scenario, the field mode solution is provided in Eq. \eqref{eq:FM-3p1-AinF}. Using this field mode solution and the de-excitation probability formula (\ref{eq:DE-gen-kasner1}), we obtain
\begin{eqnarray}\label{eq:TP-DeEx-3p1-AinF-1}
    \mathcal{P}_{\nu}^{de} (\omega) &=& \frac{4\,g^2\,\sin^2{(k_{z}z_{0})}}{(2\pi)^3\,2\,\nu}\,\bigg|\int_{-\infty}^{\infty}d\eta\,e^{a\eta/c}\,e^{i\,\nu\,t-i\,\omega\,\eta}\bigg|^2\nonumber\\
    ~&=& \frac{g^2\,c\,\omega\,\sin^2{(k_{z}z_{0})}}{2\pi^2\,\nu^{3} \,a}\,\frac{1}{1-e^{-2\pi\,c\,\omega/a}}~.
\end{eqnarray}
Like the $(1+1)$ dimensional scenario, here also one can obtain this de-excitation probability by making $\omega\to -\omega$ in the excitation probability expression of Eq. \eqref{eq:TP-3p1-AinF-1}. Therefore, in both $(1+1)$ and $(3+1)$ dimensions, if the atom follows the light cone time, the de-excitation probability can be obtained from the excitation by changing the sign of the atomic energy gap. However, this is not true when the atom is following the Minkowski time and the mirror is in the light cone region. \vspace{0.2cm}

\noindent \emph{Case(ii): Static atom at $z=z_{0}$:-$~~$}
Here, we examine the scenario where the mirror is at $\zeta = 0$ spanning from PLC to FLC, and the atom stays stationary at $z = z_{0}$, i.e., the atom is following the Minkowski time and the field modes are realized in terms of the light cone coordinates. In this case, the field mode solution, after transformation to the Minkowski coordinates, is given in Eq. \eqref{eq:FM-3p1-MinF}. Using this field mode solution along with the expression of the de-excitation probability from equation (\ref{eq:DE-gen-kasner2}), we obtain
\begin{eqnarray}\label{eq:TP-DeEx-3p1-MinF-1}
    \mathcal{P}_{\nu}^{de} (\omega) &=& g^2 \bigg|\int_{-\infty}^{\infty} dt~\phi^{\star}_{\nu,\vec{k}_{\perp}}[t,z_{0}]~e^{-i\,\omega\,t}\bigg|^2~\nonumber\\
    ~&=&\frac{g^2\,c\,\sin^2{(\omega\,z_{0}/c)}}{2\,\pi^2\,\omega^2\,a}\, \frac{1}{1-e^{-2\pi\,c\,\nu/a}}~.
\end{eqnarray}
Here also, like the $(1+1)$ dimensional case, one can see that by making either $\omega\to -\omega$ or $\nu\to -\nu$ in the expression of the excitation probability \eqref{eq:TP-3p1-MinF-2}, one cannot obtain this de-excitation probability. This observation provides additional distinction between the scenarios with the atom following the light cone and Minkowski time. Note that the de-excitation probabilities from Eqs. \eqref{eq:TP-DeEx-3p1-AinF-1} and \eqref{eq:TP-DeEx-3p1-MinF-1} are not matching when $\nu=\omega$.

\subsection{Atomic De-Excitation in Black Hole Spacetime}

In this part of the section, we investigate the de-excitation probabilities for atoms in the presence of a mirror in a Schwarzschild black hole spacetime. In particular, we consider two distinct configurations. In the first case, the atom is at $ t_{S} = 0 $ in Schwarzschild spacetime, and its proper time is associated with the Schwarzschild coordinate time $ r_{\star}/c $ inside the black hole, while the mirror is fixed at $ X = X_0 $ in Kruskal--Szekeres (KS) coordinates, and its proper time is associated with KS time. In the second case, the atom is fixed at $ X = X_0 $ in KS coordinates, and the atom's proper time is related to KS time, whereas the mirror is at $t_{S} = 0 $ in Schwarzschild coordinates, and now the mirror's proper time is related to Schwarzschild time $ r_{\star}/c $.
Subsequently, we consider the interaction Hamiltonian depicted in Eq. \eqref{eq:Int-Hamiltonian-Sch1} and follow the adjacent analysis. Then, with the help of the analysis as presented in \cite{Kumawat:2024kul}, we obtain the formula for estimating the de-excitation probability as
\begin{eqnarray}
    \mathcal{P}_{\nu}^{de} (\omega_{r_{\star}}) &=& g^2\bigg|\int_{-\infty}^{r_{\star}^{f}} \frac{dr_{\star}}{c}\,e^{\kappa_{H}\,r_{\star}}\,\phi^{\star}_{\nu}\big[t(r_{\star}),z(r_{\star})\big]\,e^{-i\,\omega_{r_{\star}}r_{\star}/c}\bigg|^2~.
    \label{eq:DE-gen-Sch1}
\end{eqnarray}
At the same time, when the atom is following the Kruskal time and the mirror is at $t_{S}=0$ in Schwarzschild coordinates, we consider the interaction Hamiltonian from Eq. \eqref{eq:Int-Hamiltonian-Sch2}, and with  the help of \cite{Kumawat:2024kul}, one can obtain the formula for de-excitation probability as
\begin{eqnarray}
    \mathcal{P}_{\nu}^{de} (\omega_{T}) &=& g^2\bigg|\int_{T_{-}}^{T_{+}} dT\,(\partial\tau/\partial T)\,\phi^{\star}_{\nu}\big[t(T),z(T)\big]\,e^{-i\,\omega_{T}T}\bigg|^2~.
    \label{eq:DE-gen-Sch2}
\end{eqnarray}
In our subsequent analysis, we shall be using Eqs. \eqref{eq:DE-gen-Sch1} and \eqref{eq:DE-gen-Sch2} to obtain the explicit expressions of the de-excitation probabilities corresponding to different scenarios.

\subsubsection{Case(i): The mirror is at $X=X_{0}$} 

Here, we examine the case where a mirror in Kruskal-Szekeres (KS) coordinates is at $X=X_{0}$, i.e., the mirror's time is measured in KS time. Meanwhile, the atom is at $t_{S}=0$ in Schwarzschild spacetime, with its time measured by $r_{\star}/c$. In this case, the field mode solution in the KS spacetime is given in Eq. \eqref{eq:FM-Sch-Kruskal}. Using this field mode solution along with the expression of the de-excitation probability of (\ref{eq:DE-gen-Sch1}), we get
\begin{eqnarray}\label{eq:TP-DeEx-Sch-MatX0}
    \mathcal{P}_{\nu}^{de} (\omega) &=& \frac{g^2}{4\,\pi\,\nu}\,\bigg|\int\,\frac{dr_{\star}}{c}\,e^{\kappa_{H}\,r_{\star}}\,e^{-i\,\omega\,r_{\star}/c}\,e^{i\,\nu\,T}\Big[e^{i\,|k|(X-X_{0})}-e^{-i\,|k|(X-X_{0})}\Big]\bigg|^2~\nonumber\\
    ~&=& \frac{2\,g^2\,\omega\,\sin^{2}(\nu\,X_{0}/c)}{\nu^{3}\,\kappa_{H}\,c}\,\frac{1}{1-e^{-2\,\pi\,\omega/(c\,\kappa_{H})}}\,\left|1-\frac{\Gamma\Big(\frac{-i\,\omega}{c\,\kappa_{H}}+1,-\frac{i\,\nu\,\chi^{f}}{c\,\kappa_{H}}\Big)}{\Gamma\Big(\frac{-i\,\omega}{c\,\kappa_{H}}+1\Big)}\right|^2~.
\end{eqnarray}
From the above expression, one can observe that this de-excitation probability can be obtained from the excitation probability of Eq. \eqref{eq:TP-Sch-MatX0-3} by making $\omega\to -\omega$ in it. This observation is similar to the ones from the light cone scenario when the atom follows the light cone time.

\subsubsection{Case(ii): The atom is at $X=X_{0}$}

Here, we look at the situation where the atom is at  $X = X_{0} $ and a mirror is at $t_{S}=0$, i.e., the atom follows KS time. In this scenario, the field mode solution in terms of the Schwarzschild coordinates and then represented in terms of the Kruskal coordinates is provided in Eq. \eqref{eq:FM-1p1-MinF-3}. Utilizing this field mode solution with the expression of the de-excitation probability of Eq. (\ref{eq:DE-gen-Sch2}), we obtain:
\begin{eqnarray}\label{eq:TP-DeEx-Sch-AatX0}
    \mathcal{P}_{\nu}^{de} (\omega) &=& \frac{g^2}{4\,\pi\,\nu}\,\Bigg|\int\,dT\,e^{-i\,\omega\,T}\,\bigg[\big\{\kappa_{H}(c\,T-X_{0})\big\}^{i\,\nu/(\kappa_{H}\,c)}\,\Theta(c\,T-X_{0})-\big\{\kappa_{H}(c\,T+X_{0})\big\}^{i\,\nu/(\kappa_{H}\,c)}\,\Theta(c\,T+X_{0})\bigg]\Bigg|^2 \nonumber\\
      ~&=& \frac{2\,g^2\,\sin^2{(\omega\,X_{0}/c)}}{\omega^2\,\kappa_{H}\,c}\,\frac{1}{1-e^{-\frac{2\,\pi\,\nu}{c\,\kappa_{H}}}}\,\big|1-\chi_{Sch}\big|^2~.
\end{eqnarray}
In the last expression $\chi_{Sch}$ is given by
\begin{eqnarray}\label{eq:TP-DeEx-Sch-AatX0-chi}
    \chi_{Sch} &\equiv& \chi_{Sch}(\omega,\,\nu)\nonumber\\
    ~&=& \frac{1}{\Gamma\left(1+\frac{i\,\nu }{c\,\kappa_{H}}\right)}\Bigg[\Gamma \left(1+\frac{i\,\nu }{c\,\kappa_{H}},\frac{i\,\omega (T_{0}-\kappa_{H}\,X_{0})}{c\,\kappa_{H}}\right)\nonumber\\
    ~&+& \frac{e^{i\,\omega\,X_{0}/c}}{2\,i\,\sin{(\omega\,X_{0}/c)}}\,\bigg\{\Gamma \left(1+\frac{i\,\nu }{c\,\kappa_{H}},\frac{i\,\omega (T_{0}+\kappa_{H}\,X_{0})}{c\,\kappa_{H}}\right)-\Gamma \left(1+\frac{i\,\nu }{c\,\kappa_{H}},\frac{i\,\omega (T_{0}-\kappa_{H}\,X_{0})}{c\,\kappa_{H}}\right)\bigg\}\Bigg]~.
\end{eqnarray}
One can notice that the above de-excitation probability cannot be obtained from the excitation probability of a similar situation from Eq. \eqref{eq:TP-Sch-AatX0-2} by making the changes of $\omega\to -\omega$ or by making $\nu \to -\nu$. Similar to the excitation probabilities, the de-excitation probabilities from Eqs. \eqref{eq:TP-DeEx-Sch-MatX0} and \eqref{eq:TP-DeEx-Sch-AatX0} are not the same when $\nu=\omega$.

\subsection{Excitation to De-excitation Ratio}

In this part of the section, we study the excitation to de-excitation ratio (EDR), which is obtained from the ratio of excitation and de-excitation probabilities, i.e., EDR is $\mathcal{R}_{\nu}(\omega) = \mathcal{P}^{ex}_{\nu}(\omega)/\mathcal{P}^{de}_{\nu}(\omega)$. We would like to point out that the individual transition probabilities (excitation as well as de-excitation) in our current scenarios are not purely Planckian in nature. In this case, the ``effective temperature'' can be accurately identified through EDR (e.g. see \cite{Barman:2021oum} and references therein). In fact, a similar proposal was also given by S. Weinberg in the case of a non-equilibrium system of photons (see e.g. the discussion in Section 6.2 of \cite{JCasas_2003}). So if temperature is a good physical parameter to check the status of quantum equivalence, then EDR can be a valuable quantity in this discussion. In addition, it is observed in \cite{Kumawat:2024kul} that EDRs can signify crucial features in regard to quantum equivalence between analogous scenarios. With this motivation, we focus on studying the EDRs in the following analysis.

In $(1+1)$ dimensions, when the atom follows light cone time and is at $\zeta=0$, and the mirror is static in Minkowski spacetime at $z=z_{0}$, we take the expressions of $\mathcal{P}^{ex}_{\nu}(\omega)$ and $\mathcal{P}^{de}_{\nu}(\omega)$ from Eqs. (\ref{eq:TP-1p1-AinF-1}) and(\ref{eq:TP-DeEx-1p1-AinF-1}), and get the expression of EDR as 
\begin{eqnarray}\label{eq:EDR-1p1-StAtomFK-StMirrorMinko}
    \mathcal{R}_{\nu}(\omega)=e^{-2\pi\,\omega\,c/a}~.
\end{eqnarray}
At the same time, in $(1+1)$ dimensions when the mirror is in light cone region, situated at $\zeta=0$, and the atom is static in Minkowski spacetime at $z=z_{0}$, we take the expressions of $\mathcal{P}^{ex}_{\nu}(\omega)$ and $\mathcal{P}^{de}_{\nu}(\omega)$ from Eqs. (\ref{eq:TP-1p1-MinF-1}) and(\ref{eq:TP-DeEx-1p1-MinF-1}), and get the expression of EDR as
\begin{eqnarray}\label{eq:EDR-1p1-StMirrorFK-StAtomMinko}
    \mathcal{R}_{\nu}(\omega)=e^{-2\pi\,\nu\,c/a}~.
\end{eqnarray}
We can see from the Eqs. (\ref{eq:EDR-1p1-StAtomFK-StMirrorMinko}) and (\ref{eq:EDR-1p1-StMirrorFK-StAtomMinko}) that when the atomic and field frequencies are equal, that is, when $\omega = \nu$, the EDR in $(1+1)$ dimensions for the static atom in Minkowski spacetime is the same as the EDR for the static atom in the light cone region.
In $(3+1)$ dimensions, when the atom is in FLC at $\zeta=0$, and the mirror is static at $z=z_{0}$, we take the expressions of $\mathcal{P}^{ex}_{\nu}(\omega)$ and $\mathcal{P}^{de}_{\nu}(\omega)$ from Eqs. (\ref{eq:TP-3p1-AinF-1}) and(\ref{eq:TP-DeEx-3p1-AinF-1}), and get the expression of EDR, which is the same as Eq. \eqref{eq:EDR-1p1-StAtomFK-StMirrorMinko}. One can also check Table \ref{tab:Obs-AllTrn} for all the expressions corresponding to the light cone regions in Minkowski spacetime.
At the same time, in $(3+1)$ dimensions when the mirror is at $\zeta=0$, and the atom is static in Minkowski spacetime at $z=z_{0}$, we take the expressions of $\mathcal{P}^{ex}_{\nu}(\omega)$ and $\mathcal{P}^{de}_{\nu}(\omega)$ from eqs. (\ref{eq:TP-3p1-MinF-1}) and(\ref{eq:TP-DeEx-3p1-MinF-1}), and get the expression of EDR same as Eq. \eqref{eq:EDR-1p1-StMirrorFK-StAtomMinko}.
These observations show that the EDR in $(3+1)$ dimensions for the static atom in Minkowski spacetime is equal to the EDR for the static atom in the light cone region when the atomic and field frequencies are equal, i.e., when $\omega = \nu$.\vspace{0.2cm}

Let us now estimate the excitation to de-excitation ratio for the black hole spacetime. When the atom is in Schwarszchild spacetime at $t_{S}=0$, and the mirror is in KS spacetime at $X=X_{0}$, we take the expressions of $\mathcal{P}^{ex}_{\nu}(\omega)$ and $\mathcal{P}^{de}_{\nu}(\omega)$ from Eqs. (\ref{eq:TP-Sch-MatX0-3}) and(\ref{eq:TP-DeEx-Sch-MatX0}), and get the expression of EDR as
\begin{eqnarray}\label{eqn:EDR-Sch-MatX0}
   \mathcal{R}_{\nu}(\omega) 
     &=& e^{-2\pi\omega/c\,\kappa_{\mathcal{H}}}\frac{\left|\Gamma\Big(\frac{i\,\omega}{c\,\kappa_{H}}+1\Big)-\Gamma\Big(\frac{i\,\omega}{c\,\kappa_{H}}+1,-\frac{i\,\nu\,\chi^{f}}{c\,\kappa_{H}}\Big)\right|^2}{\left|\Gamma\Big(\frac{-i\,\omega}{c\,\kappa_{H}}+1\Big)-\Gamma\Big(\frac{-i\,\omega}{c\,\kappa_{H}}+1,-\frac{i\,\nu\,\chi^{f}}{c\,\kappa_{H}}\Big)\right|^2}\,\nonumber\\
      &=& e^{-2\pi\omega/c\,\kappa_{\mathcal{H}}}\frac{\left|\Gamma\Big(\frac{i\,\omega}{c\,\kappa_{H}}+1\Big)-\Gamma\Big(\frac{i\,\omega}{c\,\kappa_{H}}+1,-\frac{i\,\nu\,\chi^{f}}{c\,\kappa_{H}}\Big)\right|^2}{\left|\Gamma\Big(\frac{i\,\omega}{c\,\kappa_{H}}+1\Big)-\Gamma\Big(\frac{i\,\omega }{c\,\kappa_{H}}+1,\frac{i\,\nu\,\chi^{f}}{c\,\kappa_{H}}\Big)\right|^2}~.
\end{eqnarray}
At the same time, when the mirror is in Schwarzschild spacetime at $t_{S}=0$, and the atom is in KS spacetime at $X=X_{0}$, we take the expressions of $\mathcal{P}^{ex}_{\nu}(\omega)$ and $\mathcal{P}^{de}_{\nu}(\omega)$ from Eqs. (\ref{eq:TP-Sch-AatX0-2}) and (\ref{eq:TP-DeEx-Sch-AatX0}), and get the expression of EDR as 
\begin{eqnarray}\label{eqn:EDR-Sch-AatX0-1}
    \mathcal{R}_{\nu}(\omega) = e^{-2\pi\nu/c\,\kappa_{\mathcal{H}}}~\frac{\big|1-\Psi_{Sch}\big|^2}{\big|1-\chi_{Sch}\big|^2}~.
\end{eqnarray}
Now, we consider the near-horizon limit, i.e., $r\to r_{H}$. In this limit we also have $r_{\star}\rightarrow\,-\infty$ and $T_{0}\rightarrow\,\kappa_{H}X_{0}$. Then, using this limit in the expressions of $\Psi_{Sch}$ and $\chi_{Sch}$ from Eqs. \ref{eq:TP-Sch-AatX0-Psi} and \ref{eq:TP-DeEx-Sch-AatX0-chi} we obtain
\begin{eqnarray}\label{eq:TP-Sch-AatX0-Psi2}
    \Psi_{Sch} &\simeq& \Bigg[1+ \frac{e^{-i\,\omega\,X_{0}/c}}{2\,i\,\sin{(\omega\,X_{0}/c)}}\,\Bigg\{\frac{\Gamma \left(1+\frac{i\,\nu }{c\,\kappa_{H}},-\frac{2i\,\omega \,X_{0}}{c\,}\right)}{\Gamma \left(1+\frac{i\,\nu }{c\,\kappa_{H}},\right)}-1\Bigg\}\Bigg]~, 
\end{eqnarray}
and  
\begin{eqnarray}\label{eq:TP-DeEx-Sch-AatX0-chi2}
    \chi_{Sch} &\simeq& \Bigg[1+ \frac{e^{i\,\omega\,X_{0}/c}}{2\,i\,\sin{(\omega\,X_{0}/c)}}\,\Bigg\{\frac{\Gamma \left(1+\frac{i\,\nu }{c\,\kappa_{H}},\frac{2i\,\omega \,X_{0}}{c\,}\right)}{\Gamma \left(1+\frac{i\,\nu }{c\,\kappa_{H}},\right)}-1\Bigg\}\Bigg]~.   
\end{eqnarray}
Substituting these asymptotic expressions of $\Psi_{Sch}$ and $\chi_{Sch}$ from Eqs. (\ref{eq:TP-Sch-AatX0-Psi2}) and (\ref{eq:TP-DeEx-Sch-AatX0-chi2}) in Eq. (\ref{eqn:EDR-Sch-AatX0-1}), and simplifying further we obtain
\begin{eqnarray}\label{eqn:EDR-Sch-AatX0}
    \mathcal{R}_{\nu}(\omega)
    &=&  e^{-2\pi\nu/c\,\kappa_{\mathcal{H}}}~\frac{\Big|\Gamma \left(1+\frac{i\,\nu }{c\,\kappa_{H}}\right)-\Gamma \left(1+\frac{i\,\nu }{c\,\kappa_{H}},-\frac{2i\,\omega \,X_{0}}{c\,}\right)\Big|^2}{\Big|\Gamma \left(1+\frac{i\,\nu }{c\,\kappa_{H}}\right)-\Gamma \left(1+\frac{i\,\nu }{c\,\kappa_{H}},\frac{2i\,\omega \,X_{0}}{c\,}\right)\Big|^2}~.
\end{eqnarray}
In the above expressions of Eqs. (\ref{eqn:EDR-Sch-MatX0}) and (\ref{eqn:EDR-Sch-AatX0}), the EDRs are obtained for the excitation and de-excitation probabilities in the near-horizon limit. One can notice that under the two conditions: $(i)$ $\omega = \nu$ and $(ii)$ $\chi^{f} = 2X_{0} \kappa_{H}$, the EDRs from Eqs. (\ref{eqn:EDR-Sch-MatX0}) and (\ref{eqn:EDR-Sch-AatX0}) become the same. One should also notice that in the near-horizon limit, where $\chi^{f} \to \infty$ maintaining the equivalence between the EDRs requires the atomic position $X_{0}$ to be very large, which can be achieved when $t_{S}\to \infty$ as is observed from Eq. \eqref{eq:SchIn-Kruskal}. We also observe that, unlike the light cone scenarios of Eqs. \eqref{eq:EDR-1p1-StAtomFK-StMirrorMinko} and \eqref{eq:EDR-1p1-StMirrorFK-StAtomMinko}, the Schwarzschild EDRs do not describe a purely Boltzmann distribution. In this regard, compare these equations with Eqs. \eqref{eqn:EDR-Sch-MatX0} and \eqref{eqn:EDR-Sch-AatX0}.

\section{Observations \& Discussion}\label{sec:discussion}

The analogy in atomic excitation probabilities between a uniformly accelerated atom in the presence of a static mirror and a static atom in the presence of a uniformly accelerated mirror establishes the notion of quantum equivalence between classically equivalent scenarios \cite{Svidzinsky:2018jkp}. It is natural to presume that a similar equivalence can be observed between a static atom in the presence of a free-falling mirror and a free-falling atom in the presence of a static mirror in a Schwarzschild black hole spacetime \cite{Svidzinsky:2018jkp}, signifying a quantum realization of the EEP. In the present work, we investigated similar equivalences in atomic transitions in the presence of a mirror when the set-ups are considered in Minkowski-light cone regions. We observed the resemblance of our set-ups to the Schwarzschild interior and investigated the atomic transitions there as well, facilitating an understanding of the quantum equivalence reminiscent of the EEP.

\begin{table}[h!]
    \centering
    \begin{tabular}{ |p{2.3cm}|p{1.7cm}|p{3.7cm}|p{3.2cm}|p{5.6cm}|  }
 \hline
 \multicolumn{5}{|c|}{A tabular list of our all observations} \\
 \hline Scenarios &
 Dimensions & Excitation/De-excitation & Atom trajectory in & Transition probabilities \large$\mathcal{P}^{ex/de}_{\nu}(\omega)$ \\
 \hline 
 \multirow{8}{*}{Flat spacetime} &
 \multirow{4}{*}{$\mathbf{(1+1)}$} & Excitation & light cone  & \large${{ \frac{2\,g^2\,c\,\omega}{a\,\nu^{3}}\,\frac{\sin^2{\left(\nu\,z_{0}/c\right)}}{e^{\frac{2\pi\,\omega\,c}{a}}-1}}}$ \\\cline{4-5}
 &  &  & Minkowski & \large${{ \frac{2\,g^2\,c}{a\,\omega^2}\,\frac{\sin^2{(\omega\,z_{0}/c)}}{e^{\frac{2\pi\,\nu\,c}{a}}-1}}}$ \\\cline{3-5}
 \multirow{2}{*}{} & \multirow{2}{*}{} & De-excitation & light cone & \large${{ \frac{2\,g^2\,c\,\omega}{a\,\nu^{3}}\,\frac{\sin^2{(\nu\,z_{0}/c)}}{1-e^{-\frac{2\pi\,\omega\,c}{a}}}}}$ \\\cline{4-5}
 &  &  & Minkowski & \large${{ \frac{2\,g^2\,c}{a\,\omega^2}\,\frac{\sin^2{(\omega\,z_{0}/c)}}{1-e^{-\frac{2\pi\,\nu\,c}{a}}}}}$ \\
 \cline{2-5}
 \multirow{2}{*}{} & \multirow{4}{*}{$\mathbf{(3+1)}$} & Excitation & light cone & \large${{ \frac{ g^{2}\,c\,\omega}{2\,\pi^{2}\,\nu^{3}\, a}~\frac{\sin^2{(k_{z}z_{0})}}{e^{\frac{2\pi\omega\,c}{a}}-1}}}$ \\\cline{4-5}
 &  &  & Minkowski & \large${{ \frac{g^2\,c}{2\pi^2\,\omega^2\,a}\, \frac{\sin^2{(\omega\,z_{0}/c)}}{e^{\frac{2\pi\,\nu\,c}{a}}-1}}}$ \\\cline{3-5}
 \multirow{2}{*}{} & \multirow{2}{*}{} & De-excitation & light cone &  \large${{ \frac{ g^{2}\,c\,\omega}{2\,\pi^{2}\,\nu^{3}\, a}~\frac{\sin^2{(k_{z}z_{0})}}{1-e^{-\frac{2\pi\omega\,c}{a}}}}}$\\\cline{4-5}
 &  &  & Minkowski & \large{${{ \frac{g^2\,c}{2\pi^2\,\omega^2\,a}\, \frac{\sin^2{(\omega\,z_{0}/c)}}{1-e^{-\frac{2\pi\,\nu\,c}{a}}}}}$}\\
 \hline
 \multirow{4}{*}{BH spacetime} &
 \multirow{4}{*}{$\mathbf{(1+1)}$} & Excitation & Schwarzschild--tortoise  & \large${ \frac{2\,g^2\,\omega\,\sin^{2}(\nu\,X_{0}/c)}{\nu^{3}\,\kappa_{H}\,c}\,\frac{\left|1-\gamma_{Sch}(\omega,\nu) \right|^2}{e^{\frac{2\,\pi\,\omega}{c\,\kappa_{H}}}-1}}$ \\\cline{4-5}
 &  &  & Kruskal--Szekeres & \large${\frac{4\,g^2\,\sin^2{(\omega\,X_{0}/c)}}{\omega^2\,\kappa_{H}\,c}\,\frac{|1-\Psi_{Sch}|^2}{e^{\frac{2\,\pi\,\nu}{c\,\kappa_{H}}}-1}}$ \\\cline{3-5}
 \multirow{2}{*}{} & \multirow{2}{*}{} & De-excitation & Schwarzschild--tortoise & \large${\frac{2\,g^2\,\omega\,\sin^{2}{\left(\nu\,X_{0}/c\right)}}{\nu^{3}\,\kappa_{H}\,c}\,\frac{\left|1-\gamma_{Sch}(-\omega,\nu) \right|^2}{1-e^{-\frac{2\,\pi\,\omega}{c\,\kappa_{H}}}}}$ \\\cline{4-5}
 &  &  & Kruskal--Szekeres & \large${\frac{2\,g^2\,\sin^2{\left(\omega\,X_{0}/c\right)}}{\omega^2\,\kappa_{H}\,c}\,\frac{|1-\chi_{Sch}|^2}{1-e^{-\frac{2\,\pi\,\nu}{c\,\kappa_{H}}}}}$ \\
 \hline
\end{tabular}
    \caption{The above table lists our key observations regarding the atomic transitions in the presence of a mirror in light cone regions of $(1+1)$ and $(3+1)$ dimensional Minkowski spacetime. We have also listed our observations corresponding to a $(1+1)$ dimensional Schwarzschild black hole spacetime. It is to be noted that we have listed both the atomic excitation and de-excitation probabilities from Secs. \ref{sec:Detector-response} and \ref{sec:de-excitation}. In the above expressions, $\gamma_{Sch}(\omega,\nu)$ is given by $\gamma_{Sch}(\omega,\nu)=\Gamma\big[1+i\,\omega/(c\,\kappa_{H}),-i\,\nu\,\chi^{f}/(c\,\kappa_{H})\big]/ \Gamma\big[1+i\,\omega/(c\,\kappa_{H})\big]$, and it is obtained from Eq. \eqref{eq:TP-Sch-MatX0-3}. At the same time, the expressions of $\Psi_{Sch}$ and $\chi_{Sch}$ are obtained from Eqs. \eqref{eq:TP-Sch-AatX0-Psi} and \eqref{eq:TP-DeEx-Sch-AatX0-chi} respectively.}
    \label{tab:Obs-AllTrn}
\end{table}

Our set-up in the Minkowski-light cone region consists of two specific scenarios. In one scenario, the mirror is kept static in the Minkowski spacetime at $z=z_{0}$, and the atom is confined in the future light cone region. In the other scenario, the atom is kept static at $z=z_{0}$, and the mirror is in the light cone region. Additionally, we considered both the $(1+1)$ and $(3+1)$ dimensional spacetime scenarios. Our key observations from the analysis of this set-up are as follows. 
\begin{itemize}
    \item We observed the transition probabilities to contain thermal factors for both these scenarios and in both $(1+1)$ and $(3+1)$ dimensions. For a static mirror at $z=z_{0}$, the thermal factor relates to atomic frequency $\omega$, while for a static atom at $z=z_{0}$, it relates to the field frequency $\nu$, see Eqs. \eqref{eq:TP-1p1-AinF-3}, \eqref{eq:TP-3p1-AinF-3} and \eqref{eq:TP-1p1-MinF-4}, \eqref{eq:TP-3p1-MinF-4}.

    \item In $(1+1)$ dimensions, the spectra for transitions corresponding to different scenarios become identical when the atom and the field frequencies are equal, i.e. when $\omega = \nu$, see Eqs. \eqref{eq:TP-1p1-AinF-3} and \eqref{eq:TP-1p1-MinF-4}. In $(3+1)$ dimensions, this similarity between the transitions is not present when $\omega = \nu$, see Eqs. \eqref{eq:TP-3p1-AinF-3}, \eqref{eq:TP-3p1-MinF-4}, and Table \ref{tab:Obs-AllTrn}. Here, corresponding to a static mirror at $z=z_{0}$, the atomic transition depends on the field-mode wave-vector $(k_{z})$ along the direction of separation between the mirror and the atom, which results in its dissimilarity with the transition spectra of an atom static at $z=z_{0}$.

    \item Therefore, the $(3+1)$ dimensional scenario provides a new insight into the issue of equivalence. It should also be noted that though the individual transition probabilities in $(3+1)$ dimensions do not match when $\nu=\omega$, the excitation to de-excitation ratios (EDRs) do match, see Eqs. (\ref{eq:EDR-1p1-StAtomFK-StMirrorMinko}) and (\ref{eq:EDR-1p1-StMirrorFK-StAtomMinko}). Therefore, the equivalence between the classical relative motions is better reflected in EDRs rather than the atomic transition probabilities.
\end{itemize}

Next, we noticed the similarity between the light cone coordinate transformations from the Minkowski and the Kruskal coordinate transformations from Schwarzschild coordinates. We investigated the characteristics of atomic transitions in the presence of a mirror inside a $(1+1)$ dimensional Schwarzschild black hole to compare the findings with those of the light cone scenarios. In one scenario, the mirror is at a fixed KS coordinate $X_{0}$, and in the other scenario, the atom is fixed at $X_{0}$. Our observations are as follows. 
\begin{itemize}
    \item We observed that the two transition probabilities corresponding to two different scenarios in Schwarzschild are not the same in the limit of $\omega\to \nu$. This is in contrast to the $(1+1)$ dimensional light cone scenario. Therefore, we end up concluding that even though the exterior of a black hole can mimic an accelerated observer at the quantum level \cite{Svidzinsky:2018jkp} in $(1+1)$ dimensions, the same is not true for the interior.

    \item We observed that the EDRs corresponding to different scenarios are the same for $\omega=\nu$ but only in the near-horizon regime and when $\chi^{f} = 2X_{0} \kappa_{H}$. A notable difference compared to the light cone case is that the EDRs corresponding to the Schwarzschild case are not described by purely Boltzmann distributions, see expressions \eqref{eqn:EDR-Sch-MatX0} and \eqref{eqn:EDR-Sch-AatX0}. Therefore, from the perspective of EDRs, the equivalence at the quantum level is satisfied between analogous scenarios in a certain background. However, the same is not true between different backgrounds, i.e., between the light cone and black hole scenarios.

    \item In all of the above cases, we observed a particular feature in the transition probabilities. We observed that for light cone region, the transition probabilities are $\mathbb{P}_{\nu}(\omega)\propto \sin^2{(\Omega\,z_{0}/c)}$, with $\Omega$ being either $\nu$ or $\omega$ depending on whether the mirror or the atom is kept static at $z=z_{0}$. At the same time, inside a Schwarzschild black hole, the transition probability is $\mathbb{P}_{\nu}(\omega)\propto \sin^2{(\Omega\,X_{0}/c)}$, where $\Omega$ corresponds to $\nu$ or $\omega$ depending on whether the mirror or atom is kept fixed at the KS coordinate $X_{0}$. One can notice that, like the uniformly accelerating scenario of \cite{Svidzinsky:2018jkp}, here also the transition probabilities are periodic with respect to the distance. However, unlike \cite{Svidzinsky:2018jkp}, here the transition probabilities can vanish periodically with changing $z_{0}$ or $X_{0}$.

    \item From the field-independent expressions of the transition probabilities, we observed that in the $(1+1)$ dimensional light cone region, the atomic excitations are infrared (IR) divergent, see Eqs. \eqref{Appeq:TP-1p1-AinF-4} and \eqref{Appeq:TP-1p1-MinF-5} and the discussions therein. At the same time, in $(3+1)$ dimensions, when the mirror is static at $z=z_{0}$, the divergence is of ultraviolet (UV) type. In this regard, see Eq. \eqref{Appeq:TP-3p1-AinF-4} and the related discussion. Whereas, in $(3+1)$ dimensions, when the atom is kept static at $z=z_{0}$, the transition probability \eqref{Appeq:TP-3p1-MinF-5} is not divergent at all. These observations also provide a significant theoretical understanding of the differences between all these scenarios.
\end{itemize}

Furthermore, we would like to mention that in all these analyses, the atom is excited due to the presence of photons in the mirror's frame. This is because the photon is described with respect to Killing time, which is different from the atom's Killing time. Since both the mirror and the atom are static with respect to spatial position, while the time coordinates of measurement are different, the atomic transition is completely due to the relative time translation. Hence, this phenomenon is called a timelike transition phenomenon \cite{Olson:2010jy, Quach:2021vzo}. Moreover, in this case, the temperature is given by parameters, like $a/c$ in the Minkowski-light cone case and $c\kappa_H$ for the BH case, which have the dimension of frequency. Since within the future light cone region and within the BH, no observer is allowed to move along $\eta=$ constant and $r=$ constant trajectories, we do not call these temperatures Unruh and Hawking temperatures, respectively.

It is speculated that with inertial atoms in the light cone region, an experimentally observable outcome can be obtained for $a/c\sim 100$ GHz, which can be obtained from the scaling of the atomic energy levels, see \cite{Olson:2010jy}. This makes it experimentally more plausible to look into the atomic transitions in the light cone region. Moreover, there are experimental set-ups that analogues a reflecting mirror \cite{Wilson:2011rsw, Lahteenmaki:2011cwo, Johansson:2009zz, Johansson:2010vqd,Barman:2023wkr}, which can be utilized along with analogue atomic systems \cite{Aspachs:2010hh, Rodriguez-Laguna:2016kri, Gooding:2020scc} to construct suitable experimental set-ups in this regard (see \cite{Barman:2024jpc} for a recent proposal). Therefore, we believe our current work is not only relevant to the fundamental understanding but also provides a significant scope for future experimental verification.
%

\begin{acknowledgments}

The authors would like to acknowledge the anonymous referee whose insightful comments helped in enhancing the quality of the paper. S.B. would like to thank the Science and Engineering Research Board (SERB), Government of India (GoI), for supporting this work through the National Post Doctoral Fellowship (N-PDF, File number: PDF/2022/000428).

\end{acknowledgments}

\appendix
\section{Mode independent transition probability}\label{Appn:TP-mode-indp}
In this section of the Appendix, we ponder upon the idea of getting a field mode-independent description of the atomic transition probability. In this regard, we consider the atomic motion described by the time $\mathbb{t}$. We further consider that in this scenario, the proper time $\tau$ and this specific time $\mathbb{t}$ are related among themselves as $d\tau/d\mathbb{t}=\mathcal{R}(\mathbb{t})$, which also denotes the redshift factor by which the Hamiltonian corresponding to the proper time should be modified. In particular, the interaction Hamiltonian corresponding to time $\mathbb{t}$ with the field operator $\Phi(x)$ integrated over all the field modes can be expressed as
\begin{eqnarray}\label{Appeq:Hint-Findep}
    \hat{H}_{I}(\mathbb{t}) &=& \hbar\,g\,\mathcal{R}(\mathbb{t})\,\Phi[x(\mathbb{t})]\,\big(e^{-i\,\omega\,\mathbb{t}}\,\hat{\sigma}+h.c.\big)~,
\end{eqnarray}
where the field operator is decomposed in terms of the ladder operators and the field modes $\phi_{k}(x)$ as
\begin{eqnarray}\label{Appeq:field-decomp}
    \Phi(x) &=& \int\,d^3k\,\big[\phi_{k}(x)\,\hat{a}_{k}+\phi_{k}^{\star}(x)\,\hat{a}_{k}^{\dagger}\big]~.
\end{eqnarray}
One can observe that the above interaction Hamiltonian of Eq. \eqref{Appeq:Hint-Findep} is similar to the interaction Hamiltonian linear in field operator widely used in Unruh-DeWitt detector models \cite{Barman:2021oum, Barman:2021bbw, Barman:2021kwg, Barman:2022utm, Barman:2022xht, Barman:2023rhd, Barman:2023aqk, K:2023oon, Barman:2024vah}. One can utilize the interaction Hamiltonian from Eq. \eqref{Appeq:Hint-Findep} and obtain the mode-independent transition probability as
\begin{eqnarray}\label{Appeq:TP-Mode-indep}
    \mathcal{P} ^{ex}&=& g^2\,\int\,d\mathbb{t}\,\int\,d\mathbb{t}'\,e^{i\omega(\mathbb{t}-\mathbb{t}')}\,\mathcal{R}^{\star}(\mathbb{t}')\mathcal{R}(\mathbb{t})\,\langle 0|\,\Phi[x'(\mathbb{t}')] \,\Phi[x(\mathbb{t})]\,|0\rangle~\nonumber\\
    ~&=& g^2\,\int\,d^3k\,\bigg|\int_{-\infty}^{\infty}\,d\mathbb{t}\,e^{i\,\omega\,\mathbb{t}}\,\mathcal{R}(\mathbb{t})\,\phi_{k}^{\star}[x(\mathbb{t})]\bigg|^2~.
\end{eqnarray}
One can observe that the transition probabilities $\mathcal{P}_{\nu}(\omega)$ from Eqs. \eqref{eq:TP-gen-kasner1}, \eqref{eq:TP-gen-kasner2}, \eqref{eq:TP-gen-Sch1}, and \eqref{eq:TP-gen-Sch2} and the mode-independent transition probability from the previous Eq. \eqref{Appeq:TP-Mode-indep} are related among themselves as $\mathcal{P}=\int\,d^3k\,\mathcal{P}_{\nu}(\omega)$. In the main body of the manuscript, we used this relation to obtain the mode-independent transition probabilities in possible scenarios.

\section{Expressions of $\mathcal{P}^{ex}(\omega)$ in light cone regions}\label{Appn:FieldIndp-Kasner}
In this section of the appendix, we consider the coupling strength $g$  between the atom and the field to be constant and estimate the field mode independent transition probability $\mathcal{P}^{ex}$ by integrating the $\mathbb{P}^{ex}_{\nu}(\omega)$ for the atom-mirror pairs set up in the light cone region.

\subsection{Static mirror at $z=z_{0}$ in $(1+1)$ dimensions}

First, we consider the situation when the mirror is static at $z=z_{0}$ in $(1+1)$ dimensions. We take the expression of $\mathbb{P}^{ex}_{\nu}$ from Eq. \eqref{eq:TP-1p1-AinF-3} and integrate it over $\nu$. We obtain the resulting expression
\begin{eqnarray}\label{Appeq:TP-1p1-AinF-4}
    \mathcal{P} ^{ex}(\omega) &=& \frac{4\,g^2\,\omega}{a}\,\frac{1}{e^{2\pi\,c\,\omega/a}-1}\,\frac{z_{0}^{2}}{c^{2}}\bigg[Ci\left(\frac{2\,z_{0}\,\nu}{c}\right)-\frac{\sin^{2}{\left(\frac{z_{0}\,\nu}{c}\right)}+\frac{z_{0}\,\nu}{c}\,\sin{\left(\frac{2\,z_{0}\,\nu}{c}\right)}}{2\,\left(\frac{z_{0}\,\nu}{c}\right)^{2}}\bigg]_{z_{0}\,\nu/c=0}^{\infty} ~.
\end{eqnarray}
In the above expression $Ci(z)$ denotes the cosine integral function. We would like to mention that in the ultraviolet limit of $\nu\to \infty$, the quantity inside the bracket in the above expression vanishes, i.e., the upper limit of the above expression is zero. At the same time, when $\nu\to 0$, the quantity inside the bracket in the above expression diverges out. Therefore, one can conclude that the above transition probability $\mathcal{P} ^{ex}(\omega)$ has infrared (IR) divergence.

\subsection{Static atom at $z=z_{0}$ in $(1+1)$ dimensions}

Second, we consider the situation when the atom is static at $z=z_{0}$ in $(1+1)$ dimensions. In this scenario, we consider the expression of $\mathbb{P}^{ex}_{\nu}$ from Eq. \eqref{eq:TP-1p1-MinF-4} and integrate over $\nu$. This integration without taking limits gives us
\begin{eqnarray}\label{Appeq:TP-1p1-MinF-5}
    \mathcal{P} ^{ex}(\omega) &=& \frac{2\,g^2\,\sin^2{(\omega\,z_{0}/c)}}{\pi\,\omega^2\,c}\,  \bigg[\ln\Big(1-e^{-2\,\pi\nu\,c/a}\Big)\bigg]_{\nu\,c/a=0}^{\infty}~.
\end{eqnarray}
It is evident that the above expression diverges when $\nu\to 0$, i.e., the transition probability $\mathcal{P} ^{ex}(\omega)$ is infrared (IR) divergent.

\subsection{Static mirror at $z=z_{0}$ in $(3+1)$ dimensions}

Third, we consider a static mirror at $z=z_{0}$ in $(3+1)$ dimensions. In this regard, we consider the expression of $\mathbb{P}^{ex}_{\nu}$ from eq. \eqref{eq:TP-3p1-AinF-3}. After integrating over $\nu$ we get the field mode independent expression of the transition probability
\begin{eqnarray}\label{Appeq:TP-3p1-AinF-4}
    \mathcal{P} ^{ex}(\omega) &=& \frac{g^2\,c}{2\pi^2\,\omega \,a}\,\frac{1}{e^{2\pi\,c\,\omega/a}-1}\,\frac{2\pi}{c^3}\int_{0}^{\infty} \nu \,d\nu\int_{0}^{\pi}d\theta\, \sin\theta \sin^2{\Big(\frac{\nu\,z_{0}\,\cos{\theta}}{c}\Big)}~\nonumber\\
    ~&=& \frac{g^2\,\omega}{\pi\,a\,c^2}\,\frac{1}{e^{2\pi\,c\,\omega/a}-1}\,\bigg[\ln{\left(\frac{z_{0}\,\nu}{c}\right)}+\frac{\sin{\left(\frac{2\,z_{0}\,\nu}{c}\right)}}{z_{0}\,\nu/c}-Ci\left(\frac{2\,z_{0}\,\nu}{c}\right)\bigg]_{z_{0}\,\nu/c=0}^{\infty}~.
\end{eqnarray}
Here also, in the above expression $Ci(z)$ denotes the cosine integral function. For $\nu\to 0$ the quantity inside the bracket in the above expression becomes finite and is given by $\big\{1-\gamma-\log \left(2\, z_{0}/c\right)\big\}$. At the same time, when $\nu\to \infty$, the expression inside the bracket diverges out, signifying that the transition probability is ultraviolet (UV) divergent. One can notice that in both the $(1+1)$ dimensional scenarios the transition probabilities were IR divergent, but in the current scenario in $(3+1)$ dimensions it is UV divergent.

\subsection{Static atom at $z=z_{0}$ in $(3+1)$ dimensions}

Fourth, we consider the scenario of a static atom at $z=z_{0}$ in $(3+1)$ dimensions. We take the relevant expression of $\mathbb{P}^{ex}_{\nu}$ from Eq. \eqref{eq:TP-3p1-MinF-4} and integrate over the variable $\nu$, which gives
\begin{eqnarray}\label{Appeq:TP-3p1-MinF-5}
    \mathcal{P} ^{ex}(\omega) &=&\frac{2\,g^2\,\sin^2{(\omega\,z_{0}/c)}}{\pi\,\omega^2\,a\,c^2}\,\frac{a^3}{4\,\pi^3\,c^3}\,\zeta(3)~.
\end{eqnarray}
In the above expression $\zeta(s)$ signifies the Riemann zeta function, and $\zeta(3)\sim 1.202$ is finite. Therefore, the above expression is finite without any divergences in the IR and UV limit. This is in contrast to the transition probability for a similar atom-mirror set-up from Eq. \eqref{Appeq:TP-1p1-MinF-5} corresponding to a $(1+1)$ dimensional scenario.

\section{Evaluation of $\mathcal{P}^{ex}_{\nu}(\omega)$ in different scenarios}\label{Appn:Pnu-all}
In this section of the appendix, we provide the explicit procedure to evaluate $\mathcal{P}^{ex}_{\nu}(\omega)$ corresponding to different scenarios. In this regard, we shall first consider the Minkowski-light cone scenarios. In particular, we shall discuss the $(1+1)$ and $(3+1)$ dimensional settings in this scenario. Then we shall study the black hole scenario. We would also like to mention that we shall provide the explicit procedure of evaluating $\mathcal{P}^{ex}_{\nu}(\omega)$ only when this procedure is cumbersome.

\subsection{Static atom at $z=z_{0}$ in $(1+1)$ dimensional Minkowski spacetime}\label{Appn:Pnu-stAtm-1p1}
When the atom is kept static at $z=z_{0}$ in $(1+1)$ dimensional Minkowski spacetime we consider the field mode as provided in Eq. \eqref{eq:FM-1p1-MinF-2} and obtain the atomic excitation probability as
\begin{eqnarray}\label{eqAppn:TP-1p1-MinF-2}
    \mathcal{P}^{ex}_{\nu} (\omega) &=& g^2 \bigg|\int_{-\infty}^{\infty} dt~\phi^{\star}_{\nu}[t,z_{0}]~e^{i\,\omega\,t}\bigg|^2~\nonumber\\
    ~&=& \frac{g^2}{4\pi\,\nu}\, \bigg|\int_{z_{0}/c}^{\infty} dt~e^{i\,\omega\,t}~\Big\{\frac{a}{c^2}(c\,t-z_{0})\Big\}^{\frac{i\,\nu\,c}{a}}-\int_{-z_{0}/c}^{\infty} dt~e^{i\,\omega\,t}~\Big\{\frac{a}{c^2}(c\,t+z_{0})\Big\}^{\frac{i\,\nu\,c}{a}}\bigg|^2~.
\end{eqnarray}
We consider a change of variables $x_{1}=c\,t-z_{0}$ and $x_{2}=c\,t+z_{0}$. In that scenario, the above integral further simplifies to
\begin{eqnarray}\label{eqAppn:TP-1p1-MinF-3}
    \mathcal{P}^{ex}_{\nu} (\omega) &=& \frac{g^2}{4\pi\,\nu\,c^2}\, \bigg|\int_{0}^{\infty} dx_{1}~e^{\frac{i\,\omega\,x_{1}}{c}-\frac{i\,\omega\,z_{0}}{c}}~x_{1}^{\frac{i\,\nu\,c}{a}}-\int_{0}^{\infty} dx_{2}~e^{\frac{i\,\omega\,x_{2}}{c}+\frac{i\,\omega\,z_{0}}{c}}~x_{2}^{\frac{i\,\nu\,c}{a}}\bigg|^2\nonumber\\
    ~&=& \frac{g^2\,\sin^2{(\omega\,z_{0}/c)}}{\pi\,\nu\,c^2}\, \bigg|\int_{0}^{\infty} dx_{1}~e^{\frac{i\,\omega\,x_{1}}{c}}~x_{1}^{\frac{i\,\nu\,c}{a}}\bigg|^2\nonumber\\
    ~&=& \frac{2\,g^2\,c\,\sin^2{(\omega\,z_{0}/c)}}{\omega^2\,a}\, \frac{1}{e^{2\pi\,c\,\nu/a}-1}~.
\end{eqnarray}

\subsection{Static atom at $z=z_{0}$ in $(3+1)$ dimensional Minkowski spacetime}\label{Appn:Pnu-stAtm-3p1}
When the atom is kept static at $z=z_{0}$ in $(1+1)$ dimensional Minkowski spacetime we consider the field mode as provided in Eq. \eqref{eq:FM-3p1-MinF} and obtain the atomic excitation probability as
\begin{eqnarray}\label{eq:TP-3p1-MinF-2}
    \mathcal{P}^{ex}_{\nu} (\omega) &=& g^2 \bigg|\int_{-\infty}^{\infty} dt~\phi^{\star}_{\nu,\vec{k}_{\perp}}[t,z_{0}]~e^{i\,\omega\,t}\bigg|^2~\nonumber\\
    ~&=&\frac{g^2\,c\,\cosech(\pi\,c\,\nu/a)}{16\pi^2\,a}\,\Big(\frac{a^2}{c^2\,\nu^2}\Big)\,\frac{1}{|\Gamma(i\,\nu\,c/a)|^2}\, \nonumber\\
    ~&\times& \bigg|\int_{z_{0}/c}^{\infty} dt~e^{i\,\omega\,t}~(c\,t-z_{0})^{\frac{i\,\nu\,c}{a}}-\int_{-z_{0}/c}^{\infty} dt~e^{i\,\omega\,t}~(c\,t+z_{0})^{\frac{i\,\nu\,c}{a}}\bigg|^2~.
\end{eqnarray}
We consider a change of variables $x_{1}=c\,t-z_{0}$ and $x_{2}=c\,t+z_{0}$, and proceed in a manner similar to the $(1+1)$ dimensional case, see Eqs. \eqref{eqAppn:TP-1p1-MinF-2} and \eqref{eqAppn:TP-1p1-MinF-3}. Then the above integral becomes
\begin{eqnarray}\label{eq:TP-3p1-MinF-3}
    \mathcal{P}^{ex}_{\nu} (\omega) &=& \frac{g^2\,c\,\sin^2{(\omega\,z_{0}/c)}}{2\,\pi^2\,\omega^2\,a}\, \frac{1}{e^{2\pi\,c\,\nu/a}-1}~.
\end{eqnarray}

\subsection{The mirror is at $X=X_{0}$ in Schwarzschild black hole spacetime}\label{Appn:Pnu-stMirr-BH}
When the mirror is at $X=X_{0}$ in Schwarzschild black hole spacetime, we consider the field mode as provided in Eq. \eqref{eq:FM-Sch-Kruskal} and obtain the atomic excitation probability as
\begin{eqnarray}\label{eq:TP-Sch-MatX0-1}
    \mathcal{P}^{ex}_{\nu} (\omega) &=& \frac{g^2}{4\,\pi\,\nu}\,\bigg|\int\,\frac{dr_{\star}}{c}\,e^{\kappa_{H}\,r_{\star}}\,e^{i\,\omega\,r_{\star}/c}\,e^{i\,\nu\,T}\Big[e^{i\,|k|(X-X_{0})}-e^{-i\,|k|(X-X_{0})}\Big]\bigg|^2~\nonumber\\
    ~&=& \frac{g^2}{4\,\pi\,\nu}\,\bigg|\int\,\frac{dr_{\star}}{c}\,e^{\kappa_{H}\,r_{\star}}\,e^{i\,\omega\,r_{\star}/c}\,\Big[e^{-i\,\nu\,X_{0}/c}\,\exp{\Big(\frac{i\,\nu}{c\,\kappa_{H}}\,e^{\kappa_{H}\,v}\Big)}-e^{i\,\nu\,X_{0}/c}\,\exp{\Big(\frac{i\,\nu}{c\,\kappa_{H}}\,e^{\kappa_{H}\,u}\Big)}\Big]\bigg|^2~.
\end{eqnarray}
In the last line of the previous equation, we have used the relation between the Kruskal coordinates $(T,\,X)$ and the advanced and retarded times $(v,\,u)$ from Eq. \eqref{eq:Sch-Kruskal-EF}. From the line element \eqref{eq:SchIn-Metric} in a Schwarzschild black hole background it is evident that inside the event horizon $r< r_{H}$ the coordinate $r_{\star}$ acts as time, while $t_{S}$ behaves as a spatial coordinate. Therefore, we consider the clock time of measurement to be $r_{\star}/c$, and we further consider that the atom at $t_{S}=0$. Then the advanced and retarded time coordinates are related to the $r_{\star}$ as $v=r_{\star}$ and $u=r_{\star}$. With these considerations, the transition probability of the atom can be estimated to be
\begin{eqnarray}\label{eq:TP-Sch-MatX0-2}
    \mathcal{P}^{ex}_{\nu} (\omega) &=& \frac{g^2}{4\,\pi\,\nu\,c^2}\,\bigg|\int_{-\infty}^{r_{\star}^{f}}\,dr_{\star}\,e^{\kappa_{H}\,r_{\star}}\,e^{i\,\omega\,r_{\star}/c}\,e^{-i\,\nu\,X_{0}/c}\,\exp{\Big(\frac{i\,\nu}{c\,\kappa_{H}}\,e^{\kappa_{H}\,r_{\star}}\Big)}\,\nonumber\\
    ~&&-\int_{-\infty}^{r_{\star}^{f}}\,dr_{\star}\,\,e^{\kappa_{H}\,r_{\star}}\,e^{i\,\omega\,r_{\star}/c}e^{i\,\nu\,X_{0}/c}\,\exp{\Big(\frac{i\,\nu}{c\,\kappa_{H}}\,e^{\kappa_{H}\,r_{\star}}\Big)}\bigg|^2~\nonumber\\
    ~&=& \frac{g^2\,\sin^{2}(\nu\,X_{0}/c)}{\pi\,\nu\,c^2} \,\bigg|\int_{-r_{\star}^{f}}^{\infty}\,dr_{\star}\,e^{-\kappa_{H}\,r_{\star}}\,e^{-i\,\omega\,r_{\star}/c}\,\exp{\Big(\frac{i\,\nu}{c\,\kappa_{H}}\,e^{-\kappa_{H}\,r_{\star}}\Big)}\bigg|^2~.
\end{eqnarray}
In the above expression, we have truncated the limit of the integral till $r_{\star}^{f}$ rather than to zero, corresponding to the Schwarzschild inner singularity, as the integral is approximated to give the proper results only in the near horizon region. In the integral of the first line in Eq. \eqref{eq:TP-Sch-MatX0-2} if we consider a change of variables $r_{\star}\to -r_{\star}$, we shall get the second line. Considering further a change of variable $\chi=e^{-\kappa_{H}\,r_{\star}}$, one finds the transition probability as
\begin{eqnarray}\label{eq:TP-Sch-MatX0-3b}
    \mathcal{P}^{ex}_{\nu} (\omega) &=& \frac{g^2\,\sin^{2}(\nu\,X_{0}/c)}{\pi\,\nu\,c^2\,\kappa^2_{H}}\,\bigg|\int_{0}^{\chi^{f}}\,d\chi~\chi^{\frac{i\,\omega}{\kappa_{H}\,c}}\,e^{\frac{i\,\nu\,\chi}{c\,\kappa_{H}}}\bigg|^2~\nonumber\\
    ~&=& \frac{2\,g^2\,\omega\,\sin^{2}(\nu\,X_{0}/c)}{\nu^{3}\,\kappa_{H}\,c}\,\frac{1}{e^{2\,\pi\,\omega/(c\,\kappa_{H})}-1}\,\left|1-\frac{\Gamma\Big(\frac{i\,\omega}{c\,\kappa_{H}}+1,-\frac{i\,\nu\,\chi^{f}}{c\,\kappa_{H}}\Big)}{\Gamma\Big(\frac{i\,\omega}{c\,\kappa_{H}}+1\Big)}\right|^2~.
\end{eqnarray}

\subsection{The atom is at $X=X_{0}$ in Schwarzschild black hole spacetime}\label{Appn:Pnu-stAtm-BH}
When the atom is at $X=X_{0}$ in Schwarzschild black hole spacetime, we consider the field mode as provided in Eq. \eqref{eq:FM-1p1-MinF-3} and obtain the atomic excitation probability as
\begin{eqnarray}\label{eq:TP-Sch-AatX0-1}
    \mathcal{P}^{ex}_{\nu} (\omega) &=& \frac{g^2}{4\,\pi\,\nu}\,\Bigg|\int\,dT\,e^{i\,\omega\,T}\,\bigg[\big\{\kappa_{H}(c\,T-X_{0})\big\}^{i\,\nu/(\kappa_{H}\,c)}\,\Theta(c\,T-X_{0})-\big\{\kappa_{H}(c\,T+X_{0})\big\}^{i\,\nu/(\kappa_{H}\,c)}\,\Theta(c\,T+X_{0})\bigg]\Bigg|^2.
\end{eqnarray}
The atom is fixed at $X=X_{0}$, and we have considered the near-horizon approximation to obtain the above expression of the transition probability. One can note that both past KS and future KS regions are bounded by singularity at $r=0$. Because we considered near-horizon expressions for the transition probability, we further restrict the integrals from the horizon to some $r_{\star}^{f}$ rather than exactly to $r=0$. Then one can see that the limits of the integration variable $T$ in the above expression are from $T_{-}=-(T_{0}/c\,\kappa_{H})$ to $T_{+}=(T_{0}/c\, \kappa_{H})$, with $T_{0}=(e^{2\kappa_{H}\, r_{\star}^{f}} + \kappa_{H}^2\, X_{0}^2)^{1/2}$. Next, we consider change of variables $\chi_{+}=\kappa_{H}(c\,T-X_{0})$ and $\chi_{-}=\kappa_{H}(c\,T+X_{0})$. Then the expression for the transition probability becomes
\begin{eqnarray}\label{eq:TP-Sch-AatX0-2b}
    \mathcal{P}^{ex}_{\nu} (\omega) &=& \frac{g^2}{4\,\pi\,\nu}\,\Bigg|\int_{X_{0}/c}^{T_{+}}\,dT\,e^{i\,\omega\,T}\,\big\{\kappa_{H}(c\,T-X_{0})\big\}^{i\,\nu/(\kappa_{H}\,c)}-\int_{-X_{0}/c}^{T_{+}}\,dT\,e^{i\,\omega\,T}\big\{\kappa_{H}(c\,T+X_{0})\big\}^{i\,\nu/(\kappa_{H}\,c)}\Bigg|^2~\nonumber\\
    ~&=& \frac{g^2}{4\,\pi\,\nu}\,\frac{1}{c^2\,\kappa_{H}^2}\Bigg|\int_{0}^{T_{0}-\kappa_{H}\,X_{0}}\,d\chi_{+}\,e^{\frac{i\,\omega\,\chi_{+}}{c\,\kappa_{H}}+\frac{i\,\omega\,X_{0}}{c}}\,\chi_{+}^{\frac{i\,\nu}{\kappa_{H}\,c}}-\int_{0}^{T_{0}+\kappa_{H}\,X_{0}}d\chi_{-}\,e^{\frac{i\,\omega\,\chi_{-}}{c\,\kappa_{H}}-\frac{i\,\omega\,X_{0}}{c}}\,\chi_{-}^{\frac{i\,\nu}{\kappa_{H}\,c}}\Bigg|^2~\nonumber\\
    ~&=& \frac{g^2}{4\,\pi\,\nu}\,\frac{1}{c^2\,\kappa_{H}^2}\Bigg|2\,i\,\sin{(\omega\,X_{0}/c)}\int_{0}^{T_{0}-\kappa_{H}\,X_{0}}\,d\chi_{+}\,e^{\frac{i\,\omega\,\chi_{+}}{c\,\kappa_{H}}}\,\chi_{+}^{\frac{i\,\nu}{\kappa_{H}\,c}}-\int_{T_{0}-\kappa_{H}\,X_{0}}^{T_{0}+\kappa_{H}\,X_{0}}d\chi_{-}\,e^{\frac{i\,\omega\,\chi_{-}}{c\,\kappa_{H}}-\frac{i\,\omega\,X_{0}}{c}}\,\chi_{-}^{\frac{i\,\nu}{\kappa_{H}\,c}}\Bigg|^2~\nonumber\\
    ~&=& \frac{2\,g^2\,\sin^2{(\omega\,X_{0}/c)}}{\omega^2\,\kappa_{H}\,c}\,\frac{1}{e^{\frac{2\,\pi\,\nu}{c\,\kappa_{H}}}-1}\,\big|1-\Psi_{Sch}\big|^2~,
\end{eqnarray}
where
\begin{eqnarray}\label{eq:TP-Sch-AatX0-Psi-b}
    \Psi_{Sch} &\equiv& \Psi_{Sch}(\omega,\,\nu)\nonumber\\
    ~&=& \frac{1}{\Gamma\left(1+\frac{i\,\nu }{c\,\kappa_{H}}\right)}\Bigg[\Gamma \left(1+\frac{i\,\nu }{c\,\kappa_{H}},-\frac{i\,\omega (T_{0}-\kappa_{H}\,X_{0})}{c\,\kappa_{H}}\right)\nonumber\\
    ~&+& \frac{e^{-i\,\omega\,X_{0}/c}}{2\,i\,\sin{(\omega\,X_{0}/c)}}\,\bigg\{\Gamma \left(1+\frac{i\,\nu }{c\,\kappa_{H}},-\frac{i\,\omega (T_{0}+\kappa_{H}\,X_{0})}{c\,\kappa_{H}}\right)-\Gamma \left(1+\frac{i\,\nu }{c\,\kappa_{H}},-\frac{i\,\omega (T_{0}-\kappa_{H}\,X_{0})}{c\,\kappa_{H}}\right)\bigg\}\Bigg]~.
\end{eqnarray}

\section{Field mode solutions in region $FLC$ in $(3+1)$ dimensions}\label{Appn:Mode-3p1-F}
In this section of the Appendix, we evaluate the normalized field mode solutions in the future light cone region, i.e., in the region $FLC$ see Fig. \ref{fig:SD-Kasner}, in a $(3+1)$ dimensional Minkowski spacetime. The line element in terms of the coordinates of Eq. \eqref{eq:CT-F} in $FLC$ is given by $ds^2=e^{2\,a\,\eta/c}(-c^2\,d\eta^2+d\zeta^2)+dx^2+dy^2$. For a massless minimally coupled scalar field $\Phi$ the equation of motion $\partial_{\mu}\big(\sqrt{-g}g^{\mu\nu}\partial_{\nu}\Phi\big) = 0$, in the region $FLC$ becomes 
\begin{eqnarray}\label{Appeq:FM-EOM-1}
    -\frac{\partial_{\eta}^2\Phi}{c}+c\,\partial_{\zeta}^2\Phi+c\,e^{2\,a\,\eta/c}\,\Big(\partial_{x}^2\Phi+\partial_{y}^2\Phi\Big) &=& 0~.
\end{eqnarray}
From the line element is evident that the metric components are independent of the coordinates $(\zeta,\, x,\, y)$. Thus the field mode will be plane wave-like with respect to these coordinates, which is also apparent from the equation of motion of Eq. \eqref{Appeq:FM-EOM-1}. This motivates us to consider a field mode solution $\phi_{\nu,k_{\perp}}\sim R(\eta)\, e^{i\,(\nu\zeta/c)+i\,\Vec{k}_{\perp}.\Vec{x}_{\perp}}$. Substituting this decomposition in the above equation of motion \eqref{Appeq:FM-EOM-1} gives 
\begin{eqnarray}\label{Appn:FM-EOM-2}
    \partial_{\eta}^2R+(\nu^2+c^2\,k_{\perp}^2\,e^{2\,a\,\eta/c})=0~. 
\end{eqnarray}
One can obtain the solution of this equation in terms of the Bessel function of the first kind. Finally one can express the normalized mode functions as 
\begin{eqnarray}\label{Appeq:FM-3p1-AinF-gen}
    \phi_{\nu,k_{\perp}} &=& \sqrt{\frac{c\,\cosech(\pi\,c\,\nu/a)}{16\pi^2\,a}}\,\mathrm{J}_{-\frac{i\,c\,\nu}{a}}\bigg(\frac{c^2k_{\perp}\,e^{a\,\eta/c}}{a}\bigg) \,e^{\frac{i\,\nu\,\zeta}{c}+i\,\Vec{k}_{\perp}.\Vec{x}_{\perp}}~,
\end{eqnarray}
where $\mathrm{J}_{n}(x)$ denotes the Bessel function of first kind of order $n$. In the previous expression $\vec{x}_{\perp}=(x,\,y)$, denote the transverse directions. We have used the expression of the mode function from Eq. \eqref{Appeq:FM-3p1-AinF-gen} in investigating the transition probability of two-level atoms.

\bibliographystyle{utphys1.bst}

\bibliography{bibtexfile}

\end{document}